\documentclass[aps,pra,twocolumn,superscriptaddress,nofootinbib,floatfix]{revtex4-2}

\usepackage{amsmath,amssymb,amsfonts,bm}
\usepackage{graphicx}
\usepackage{booktabs}
\usepackage{xcolor}
\usepackage{etoolbox}
\usepackage[colorlinks=true,citecolor=blue,linkcolor=black,urlcolor=blue]{hyperref}


\newcommand{\ii}{\mathrm{i}}
\newcommand{\ee}{\mathrm{e}}
\newcommand{\ps}{\mathrm{PS}}
\newcommand{\pa}{\mathrm{PA}}
\newcommand{\pc}{\mathrm{PC}}
\newcommand{\sv}{\mathrm{SV}}
\newcommand{\ket}[1]{\lvert #1\rangle}
\newcommand{\bra}[1]{\langle #1\rvert}

\begin{document}

\title{Heralded Non-Gaussian Squeezed-State Inputs for Parity-Detection SU(1,1) Interferometry}

\author{Lifen Guo}
\affiliation{School of Physics, Jiangxi Normal University, Nanchang, China}
\author{Qingqian Kang}
\affiliation{School of Photoelectric Engineering, Jiangxi Modern Polytechnic College, Nanchang, China}
\author{Teng Zhao}
\affiliation{School of Physics, Jiangxi Normal University, Nanchang, China}
\author{Cunjin Liu}
\affiliation{School of Physics, Jiangxi Normal University, Nanchang, China}
\author{Xin Su}
\affiliation{School of Physics, Jiangxi Normal University, Nanchang, China}
\author{Liyun Hu}
\email{hlyun@jxnu.edu.cn}
\affiliation{School of Physics, Jiangxi Normal University, Nanchang, China}

\begin{abstract}
Non-Gaussian operations can reshape the photon statistics of continuous-variable probes, but their metrological advantage is meaningful only when heralding probability and photon-number resources are counted consistently. We compare photon subtraction, photon addition, and photon catalysis as input-side heralding operations in a balanced SU(1,1) interferometer with parity detection. A unified finite-transmissivity map supplies closed conditional moments and the corresponding quantum Fisher information at arbitrary operation order; internal loss is absorbed into a single effective parity observable whose lossless limit recovers the ideal pulled-back measurement. At fixed preparation parameters, single-photon subtraction and addition improve the conditional phase information over the Gaussian reference across most of the high-transmissivity regime, while multi-photon catalysis opens useful low-transmissivity windows. However, when the coherent--squeezed allocation is independently optimized at fixed conditional-probe energy and fixed interferometer gain, the success-weighted Fisher information of all three non-Gaussian operations remains below the optimized Gaussian benchmark. This conclusion is subject to the tested constraints: single-photon operations, a coherent-plus-squeezed-vacuum Gaussian family, fixed gain, and parity readout. Photon catalysis separately generates a conditional branch with high local quantum Fisher information that dark-point parity extracts poorly, identifying a measurement mismatch rather than a state-preparation failure. The result draws a sharp boundary between conditional non-Gaussian enhancement and practically available precision under explicitly stated resource constraints.
\end{abstract}

\maketitle

\section{Introduction}

Quantum-enhanced interferometry uses nonclassical states and measurements to improve phase estimation \cite{Caves1981,Braunstein1994,Paris2009,BraunsteinVanLoock2005,Demkowicz2015,Pezze2018}. In an SU(1,1) interferometer, optical parametric amplifiers replace the passive beam splitters of a Mach--Zehnder interferometer \cite{Yurke1986}. The first amplifier creates phase-sensitive two-mode correlations, and the second reverses the amplification and converts the encoded phase into an output signal. This active structure supports coherent and squeezed inputs, parity or homodyne detection, and explicit treatments of optical loss \cite{Plick2010,Ou2012,Marino2012,Chekhova2016,OuLi2020,Szigeti2017,Linnemann2016,Li2016ParitySU11,Gao2016Lossy}. It has also been realized in several experimental and multimode settings \cite{Jing2011,Kong2013,Hudelist2014,Manceau2017,Anderson2017,Gupta2018,Frascella2019,Hong2024Distributed}.

Photon subtraction (PS), photon addition (PA), and photon catalysis (PC) convert a Gaussian state into conditional non-Gaussian states with altered photon-number statistics and pair coherence. Their beam-splitter and heralding implementations are well established in optical state engineering \cite{Agarwal1991,Dakna1997,Zavatta2004,Lvovsky2002,Parigi2007}. Finite-transmissivity operations, parity detection, and success probabilities have been studied in passive Mach--Zehnder interferometers \cite{Kumar2022,Wang2018PhotonAdded,Zhao2024Catalysis}. Related studies of SU(1,1) interferometers have considered photon-operated inputs \cite{Gong2016Intramode,Guo2018PhotonAddedSU11,Xin2021PhotonLevelSU11,Hou2023ParityPS}, loss \cite{Xu2023PhotonOps,Kang2024MultiPS}, subtraction at an output port \cite{Jiang2024OutputPS}, delocalized internal subtraction \cite{Li2025DelocalizedPS}, and lossy homodyne detection for general inputs \cite{Jana2026Arbitrary}. Because these works use different operation positions, measurements, and resource conventions, their reported advantages are not directly comparable.

The position of the conditional operation is important. An operation before the first optical parametric amplifier changes the input that is subsequently amplified, whereas an internal operation acts on an already correlated two-mode state. These two arrangements are not equivalent because photon operations do not commute with two-mode squeezing \cite{Xu2023PhotonOps,Kang2024MultiPS}. We study only input-side preparation, so the non-Gaussian state is fully defined before it enters the interferometer.

We inject a coherent state into mode $a$ and a heralded non-Gaussian squeezed-vacuum state into mode $b$. The preparation is described by one finite-transmissivity map whose three choices give PS, PA, and PC. The map determines the normalized conditional state, its preparation probability, and the moments needed for phase estimation. Beyond collecting the three operations in one notation, the present analysis connects their arbitrary-order finite-transmissivity moments to a common QFI formula, identifies operation-dependent transmissivity windows from moment-based differences, and pulls internal loss back to a single effective parity observable. These three steps permit conditional, readout-limited, and per-attempt comparisons to be made without changing the underlying state model.

Our analysis follows the physical model: conditional operations and input moments are derived first, followed by the QFI and parity-based phase sensitivity. Parameter dependence is examined only after the analytic relations are established. Improvement scans use differences from the Gaussian reference; dimensionless ratios serve as diagnostics, and $P_jF_j$ quantifies information per ideal module attempt. Internal loss enters through an equivalent parity observable, with the lossless limit checked before numerical evaluation.

The paper is organized as follows. Section~II introduces the Gaussian input, unified heralding map, input moments, and photon-number resources. Section~III treats ideal conditional phase estimation (QFI and parity detection) followed by difference-based comparison and resource constraints. Section~IV derives the effective parity observable under internal loss and presents representative comparisons at selected internal transmissions. Sections~V and VI discuss the results and conclude. Technical derivations are collected in the appendices.

\section{Non-Gaussian input preparation and SU(1,1) interferometric model}

\begin{figure*}[t]
\includegraphics[width=0.70\textwidth]{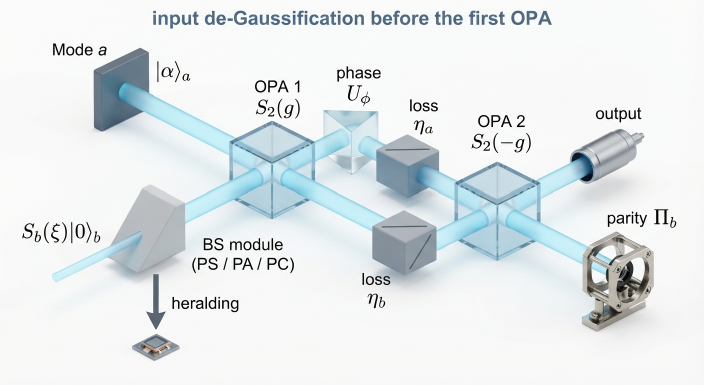}
\caption{SU(1,1) interferometer with input-side non-Gaussian state preparation. A coherent state enters mode $a$. The squeezed vacuum in mode $b$ is mixed with ancillary mode $c$, prepared in a Fock state, in a heralded beam-splitter module. The choices $(\mu,\nu)=(0,m)$, $(m,0)$, and $(m,m)$ give $m$-photon subtraction, addition, and catalysis, respectively. The conditional product input enters the first optical parametric amplifier, acquires a phase shift in mode $a$, and is recombined by the inverse amplifier. Section~IV introduces internal transmissions $\eta_a$ and $\eta_b$ before the output parity measurement on mode $b$.}
\label{fig:schematic}
\end{figure*}

We consider the preparation and interferometer shown in Fig.~\ref{fig:schematic}. A coherent state enters mode $a$. In mode $b$, a squeezed vacuum is mixed with an ancillary Fock state in a beam-splitter module, and a photon-number measurement on the ancillary output heralds the desired non-Gaussian state \cite{Dakna1997,Lvovsky2002,Parigi2007}. The two-mode product state then enters the first optical parametric amplifier (OPA), after which a phase is encoded in mode $a$. The second amplifier applies the inverse transformation, and parity is measured at the output of mode $b$.

\subsection{Gaussian input states}

We use units with $\hbar=1$. Mode $a$ is prepared in a coherent state
$\ket{\alpha}_a$, defined by $a\ket{\alpha}_a=\alpha\ket{\alpha}_a$. Mode $b$
contains the squeezed vacuum
\begin{equation}
  \ket{\xi}=S_b(\xi)\ket{0},\qquad
  S_b(\xi)=\exp\left[\frac{1}{2}\left(\xi^\ast b^2-\xi b^{\dagger 2}\right)\right],
\end{equation}
with $\xi=r\ee^{\ii\theta_s}$. Its Fock expansion is
\begin{equation}
  \ket{\xi}
  =
  \frac{1}{\sqrt{\cosh r}}
  \sum_{\ell=0}^{\infty}
  \frac{\sqrt{(2\ell)!}}{2^\ell \ell!}
  \left[-\ee^{\ii\theta_s}\tanh r\right]^\ell
  \ket{2\ell}.
  \label{eq:sv_coeff}
\end{equation}
Only even photon numbers occur. We write
$\langle\hat O\rangle_{\sv}=\bra{\xi}\hat O\ket{\xi}$ for an expectation value
in mode $b$ and define $\hat n_b=b^\dagger b$. The three moments required below are
\begin{equation}
  \begin{aligned}
  N_{\sv}&\equiv\langle\hat n_b\rangle_{\sv}=\sinh^2r,\\
  V_{\sv}&\equiv\langle\hat n_b^2\rangle_{\sv}-N_{\sv}^2
  =2\sinh^2r\cosh^2r,\\
  M_{\sv}&\equiv\langle b^2\rangle_{\sv}
  =-\ee^{\ii\theta_s}\sinh r\cosh r.
  \end{aligned}
  \label{eq:sv_moments}
\end{equation}
Here $N_{\sv}$ is the mean photon number, $V_{\sv}$ is the photon-number
variance, and $M_{\sv}$ is the pair-coherence moment. The squeezing phase
$\theta_s$ fixes the phase of $M_{\sv}$ and
therefore its interference with the coherent input in the phase-information
formula derived in Sec.~\ref{sec:qfi}.

\subsection{Heralded preparation and unified map}

The ancillary mode $c$ is prepared in the Fock state $\ket{\mu}_c$, mixed with
mode $b$, and projected onto $\ket{\nu}_c$ at the output. Here
$\mu,\nu\in\mathbb N_0$ are the incident and detected ancilla photon numbers,
respectively, so that $\hat n_c\ket{\mu}_c=\mu\ket{\mu}_c$ and
$\hat n_c\ket{\nu}_c=\nu\ket{\nu}_c$. A successful outcome applies the conditional operator
${}_c\!\bra{\nu}\hat B_{bc}(T)\ket{\mu}_c$ to the squeezed vacuum. The operator
$\hat B_{bc}(T)$ is the unitary beam-splitter transformation that mixes modes $b$
and $c$, as in conditional optical state engineering \cite{Dakna1997,Lvovsky2002}.
In all three operations, $T$ denotes the intensity transmissivity of
the corresponding heralding module.

For intensity transmissivity $0<T\leq1$, define $R=1-T$. We use the
beam-splitter convention
\begin{align*}
 \hat B_{bc}(T)b^\dagger\hat B_{bc}^\dagger(T)
 &=\sqrt{T}\,b^\dagger+\sqrt{R}\,c^\dagger,\\
 \hat B_{bc}(T)c^\dagger\hat B_{bc}^\dagger(T)
 &=-\sqrt{R}\,b^\dagger+\sqrt{T}\,c^\dagger.
\end{align*}
Introduce two auxiliary variables through
\[
 \ket{\mu}_c=
 \left.\frac{\partial_u^\mu}{\sqrt{\mu!}}
 \ee^{u c^\dagger}\ket{0}_c\right|_{u=0},\qquad
 {}_c\!\bra{\nu}=
 \left.\frac{\partial_v^\nu}{\sqrt{\nu!}}
 {}_c\!\bra{0}\ee^{v c}\right|_{v=0}.
\]
The ancilla matrix element and its operator-valued kernel are
\begin{equation}
  \begin{aligned}
  \hat K_{\mu,\nu}(T)
  &= {}_c\!\bra{\nu}\hat B_{bc}(T)\ket{\mu}_c\\
  &=\left.\frac{\partial_u^\mu\partial_v^\nu}
  {\sqrt{\mu!\nu!}}\,
  \mathcal K(u,v;T)\right|_{u=v=0},\\
  \mathcal K(u,v;T)
  &=\ee^{-\sqrt{R}\,u b^\dagger}\ee^{\sqrt{T}\,uv}
    \ee^{\sqrt{R/T}\,vb}\,T^{\hat n_b/2}.
  \end{aligned}
  \label{eq:unified_kraus}
\end{equation}
The factors in $\mathcal K$ act in the displayed order.  Both ancilla
Fock states are generated by finite differentiation, and
Eq.~\eqref{eq:unified_kraus} gives the conditional operator itself rather than
only its number-basis coefficients.
For a signal Fock component $\ket{\ell}_b$ with $\ell+\mu-\nu\ge0$, this map has the form
\begin{equation}
  \hat K_{\mu,\nu}(T)\ket{\ell}_b
  =C_{\ell}^{(\mu,\nu)}(T)\ket{\ell+\mu-\nu}_b,
  \label{eq:unified_kraus_number}
\end{equation}
where
\begin{align}
 \mathcal G_{\ell}^{(\mu)}(x;T)
 &=(\sqrt{T}+\sqrt{R}\,x)^\ell
   (-\sqrt{R}+\sqrt{T}\,x)^\mu,\nonumber\\
 C_{\ell}^{(\mu,\nu)}(T)
  &=\sqrt{\frac{\nu!(\ell+\mu-\nu)!}{\ell!\mu!}}\,
  \left[x^\nu\right]\mathcal G_{\ell}^{(\mu)}(x;T).
  \label{eq:unified_kraus_coefficient}
\end{align}
Here $x$ is a formal auxiliary variable, and $[x^\nu]f(x)$ denotes the
coefficient of $x^\nu$ in $f(x)$. This coefficient extraction is the
number-basis form of the double derivative in Eq.~\eqref{eq:unified_kraus}.
The success probability and normalized
conditional state are
\begin{equation}
  \begin{aligned}
  P_{\mu,\nu}&=\bra{\xi}\hat K_{\mu,\nu}^{\dagger}(T)\hat K_{\mu,\nu}(T)\ket{\xi},\\
  \rho_{\mu,\nu}&=\frac{\hat K_{\mu,\nu}(T)\ket{\xi}\bra{\xi}\hat K_{\mu,\nu}^{\dagger}(T)}{P_{\mu,\nu}}.
  \end{aligned}
  \label{eq:unified_conditional_state}
\end{equation}

The three operations are special cases of the same map. PS, PA, and PC
correspond to $(\mu,\nu)=(0,m)$, $(m,0)$, and $(m,m)$,
respectively, with $m\geq1$. For $j\in\{\ps,\pa,\pc\}$, we write the
corresponding operator, probability, and state as $\hat K_j^{(m)}$,
$P_j^{(m)}$, and $\rho_j^{(m)}$. The normalized two-mode state immediately
before the first optical parametric amplifier is
\begin{equation}
  \ket{\psi_{\rm in}^{(j,m)}}=
  \ket{\alpha}_a\otimes\ket{\chi_j^{(m)}}_b,
  \qquad
  \ket{\chi_j^{(m)}}_b=
  \frac{\hat K_j^{(m)}S_b(\xi)\ket{0}_b}{\sqrt{P_j^{(m)}}}.
  \label{eq:input}
\end{equation}
For any single-mode observable $\hat O$, the order-dependent conditional moment follows directly from the finite-transmissivity operator,
\begin{equation}
  \langle\hat O\rangle_j^{(m)}=
  \frac{\bra{\xi}\hat K_j^{(m)\dagger}\hat O\hat K_j^{(m)}\ket{\xi}}{P_j^{(m)}}.
  \label{eq:general_m_moments}
\end{equation}
Equations~\eqref{eq:unified_kraus_coefficient} and \eqref{eq:general_m_moments} define the general-$m$ moments at the operator level. Appendix~\ref{app:moments} expresses them as finite differential operators acting on the closed squeezed-vacuum generating function $G(z)=(1-z)^{-1/2}$, including the pair-coherence moment for PS, PA, and PC. For every fixed photon number $m$, the analytic derivatives terminate at finite order and require no Fock-space truncation. A superscript $(m)$ is omitted when one fixed order is considered and restored when different orders are compared. Replacing $\hat K_{\ps}$ by $b^m$ or $\hat K_{\pa}$ by $b^{\dagger m}$ would remove the heralding filter and its success probability, so the finite-transmissivity conditional operators are retained throughout.

PS is implemented by mixing the signal with vacuum at a beam splitter of intensity transmissivity $T$ and detecting $m$ photons in the ancillary output. Its Kraus operator is
\begin{equation}
  \hat K_{\ps}^{(m)}(T)
  =
  \frac{R^{m/2}}{\sqrt{m!}}\,
  T^{\hat n_b/2} b^m .
  \label{eq:kps}
\end{equation}
In the number basis,
\begin{equation}
\begin{aligned}
  \hat K_{\ps}^{(m)}(T)\ket{n}
  &=
  \begin{cases}
  \kappa_{n,m}^{(\ps)}(T)\ket{n-m},& n\ge m,\\
  0,&n<m,
  \end{cases}\\
  \kappa_{n,m}^{(\ps)}(T)
  &=
  \sqrt{\binom{n}{m}}\,
  R^{m/2}T^{(n-m)/2}.
\end{aligned}
  \label{eq:kps_number}
\end{equation}

PA is implemented with an $m$-photon ancilla. The signal mode $b$ and the ancillary mode $c$ are mixed at a beam splitter of intensity transmissivity $T$, and the ancillary output is projected onto vacuum. Up to an irrelevant phase, the corresponding Kraus operator is
\begin{equation}
  \hat K_{\pa}^{(m)}(T)
  =
  \frac{R^{m/2}}{\sqrt{m!}}\,
  b^{\dagger m}T^{\hat n_b/2},
  \label{eq:kpa}
\end{equation}
so that
\begin{equation}
  \hat K_{\pa}^{(m)}(T)\ket{n}
  =
  \sqrt{\binom{n+m}{m}}\,
  R^{m/2}T^{n/2}\ket{n+m}.
  \label{eq:kpa_number}
\end{equation}
Thus PA is treated on the same linear-optical footing as PS and PC, although it requires a nonclassical $m$-photon ancilla.

PC mixes the signal with an $m$-photon ancilla and postselects
the same photon number at the ancillary output. Setting $\mu=\nu=m$ in
Eq.~\eqref{eq:unified_kraus} gives the equivalent differential and
normal-ordered forms
\begin{equation}
  \begin{aligned}
  \hat K_{\pc}^{(m)}(T)
  &=\left.\frac{1}{m!}\partial_u^m\partial_v^m
    \mathcal K(u,v;T)\right|_{u=v=0}\\
  &=T^{m/2}:\!L_m\!\left(
    \frac{R}{T}\hat n_b\right)\!:
    T^{\hat n_b/2},
  \end{aligned}
  \label{eq:kpc}
\end{equation}
where $L_m$ is the Laguerre polynomial and $:\cdots:$ denotes normal
ordering.  Equation~\eqref{eq:kpc} makes explicit that PC is a number-diagonal finite polynomial followed by
attenuation.
Its number-basis action is
\begin{equation}
  \hat K_{\pc}^{(m)}(T)\ket{n}
  =
  A_n^{(m)}(T)\ket{n},
  \label{eq:kpc_number}
\end{equation}
where
\begin{equation}
\begin{aligned}
  A_n^{(m)}(T)=
  \sum_{q=0}^{\min(n,m)}
  (-1)^q
  \binom{n}{q}\binom{m}{q}
  &\\
  {}\times
  T^{(n+m-2q)/2}R^q .
\end{aligned}
  \label{eq:cat_amp}
\end{equation}
Equation~\eqref{eq:cat_amp} follows from
$L_m(x)=\sum_{q=0}^m\binom{m}{q}(-x)^q/q!$ and
$\langle n|:\!\hat n_b^q\!:|n\rangle=q!\binom{n}{q}$.  Equation~\eqref{eq:cat_amp} therefore follows from the
number-state expansion of the differential operator in
Eq.~\eqref{eq:kpc}.
For $m=1$, $A_n^{(1)}=T^{(n-1)/2}(T-nR)$, which shows explicitly that PC changes the photon-number weights without changing photon-number parity.

The fully transmitting boundary differs among the three operations. At $T=1$,
the PS and PA Kraus operators vanish, and their success probabilities are zero.
The familiar ideal operators $b^m$ and $b^{\dagger m}$ describe only the
normalized conditional limit $T\to1^{-}$ after the common
vanishing factor has been removed. In contrast, PC gives
$\hat K_{\pc}^{(m)}(1)=I$ and $P_{\pc}^{(m)}(1)=1$. Therefore, $T=1$ is a
zero-probability boundary for PS and PA, but an identity operation for PC.

\subsection{Operation-order-dependent input moments}

Let $j\in\{\ps,\pa,\pc\}$ label the heralded operation and let
$m=1,2,\ldots$ be the number of photons subtracted, added, or returned in the
PC heralding event.  We use
$\langle\hat O\rangle_j^{(m)}=\operatorname{Tr}[\rho_j^{(m)}\hat O]$.
The three quantities entering the ideal QFI are
\begin{align}
  N_j^{(m)}&=\langle \hat n_b\rangle_j^{(m)},\\
  V_j^{(m)}&=\langle \hat n_b^2\rangle_j^{(m)}-\langle \hat n_b\rangle_j^{(m)2},\\
  M_j^{(m)}&=\langle b^2\rangle_j^{(m)}.
  \label{eq:moments}
\end{align}
Here $N_j^{(m)}$ is the conditional mean photon number in mode $b$,
$V_j^{(m)}$ is its number variance, and $M_j^{(m)}$ is the
pair-coherence moment. The Gaussian baseline uses
$N_{\sv}$, $V_{\sv}$, and $M_{\sv}$ from Eq.~\eqref{eq:sv_moments} and carries no operation
index $m$. Together with $P_j^{(m)}$, these moments specify how the operation
order enters the subsequent phase-estimation formulas. Because the squeezed vacuum
has even photon-number parity, PS and PA have conditional parity $(-1)^m$,
while PC preserves even parity. Hence every normalized conditional state
$\ket{\chi_j^{(m)}}$ is a pure parity eigenstate, and all odd-parity
moments vanish, including
\begin{equation}
  \langle b\rangle_j^{(m)}=0,
  \qquad
  \langle\{\hat n_b,b\}\rangle_j^{(m)}
  =
  \langle\{\hat n_b,b^\dagger\}\rangle_j^{(m)}=0.
  \label{eq:zero_first_moment}
\end{equation}
The number-weighted identities, not the zero first moment alone, reduce the general pure-state QFI to the compact three-moment formula in Sec.~\ref{sec:qfi}. In the single-order derivations below, $N_j$, $V_j$, and $M_j$ denote the corresponding fixed-$m$ values.

In the ideal operator limit, single-photon PS and PA generate the same normalized squeezed single-photon state:
\begin{equation}
  bS(\xi)\ket{0}\propto S(\xi)\ket{1},\qquad
  b^\dagger S(\xi)\ket{0}\propto S(\xi)\ket{1}.
  \label{eq:one_photon_equiv}
\end{equation}
At the same transmissivity $T$ and with matched phase conventions, the
finite-transmissivity subtraction and addition modules produce the same
conditional state and moments; the common attenuation replaces the squeezing
parameter by an effective value. PS and PA can therefore have identical
conditional moments while having different success probabilities. This exact
normalized-state equivalence is special to $m=1$; for higher order the finite
differential generators in Appendix~\ref{app:moments} contain different PS and
PA polynomials. The single-photon case therefore illustrates an important point: an
operation ranking based only on normalized conditional states can be
experimentally misleading.

\subsection{Analytic moments for single-photon operations}
\label{sec:analytic_moments}

The phase-estimation formulas require only the moments in
Eq.~\eqref{eq:moments}. Equations~\eqref{eq:unified_kraus_coefficient} and
\eqref{eq:general_m_moments} generate them at arbitrary $m$ by finite
differentiation. We now give the closed single-photon results. The squeezing
phase is chosen so that the pair-coherence moment is real and positive; its
general phase can be restored at the end.

For single-photon PS and PA, Eqs.~\eqref{eq:kps} and \eqref{eq:kpa} give the
same normalized conditional photon-number distribution,
\begin{equation}
  \begin{aligned}
  p_l^{\ps}=p_l^{\pa}
  &=(1-z)^{3/2}(2l+1)\binom{2l}{l}\frac{z^l}{4^l},\\
  z&=T^2\tanh^2r,\qquad n_b=2l+1 .
  \end{aligned}
  \label{eq:odd_distribution}
\end{equation}
The equality follows by applying the two Kraus operators to the
squeezed-vacuum coefficients and relabeling the remaining odd photon number.
The normalization and moments follow from
\begin{equation}
  \sum_{l=0}^{\infty}(2l+1)\binom{2l}{l}\frac{z^l}{4^l}
  =
  \frac{1}{(1-z)^{3/2}},
  \label{eq:odd_generating}
\end{equation}
and its derivatives with respect to $z$:
\begin{align}
  N_{\ps}=N_{\pa}
  &=
  \frac{1+2z}{1-z},\\
  V_{\ps}=V_{\pa}
  &=
  \frac{6z}{(1-z)^2},\\
  M_{\ps}=M_{\pa}
  &=
  \frac{3\sqrt{z}}{1-z}.
  \label{eq:ps_pa_moments}
\end{align}
The corresponding success probabilities are different:
\begin{align}
  P_{\pa}
  &=
  \frac{1-T}{\cosh r}\frac{1}{(1-z)^{3/2}},\\
  P_{\ps}
  &=
  T\tanh^2r\,P_{\pa}.
  \label{eq:ps_pa_success}
\end{align}
At equal transmissivity, single-photon PS and PA have identical normalized
conditional moments. Their preparation probabilities, however, are different.
Their conditional phase-estimation formulas therefore agree,
whereas any probability-weighted quantity distinguishes them.

For single-photon PC, the state remains even. Its success probability and
photon-number moments are
\begin{equation}
  \begin{aligned}
  P_{\pc}&=\frac{Z_{\pc}}{T\cosh r},\\
  N_{\pc}&=2\mathcal D\ln Z_{\pc},\qquad
  V_{\pc}=4\mathcal D^2\ln Z_{\pc}.
  \end{aligned}
  \label{eq:pc_NV}
\end{equation}
where the normalization function is
\begin{equation}
  \begin{aligned}
  Z_{\pc}&=\left(T-2R\mathcal D\right)^2G(z),\\
  G(z)&=(1-z)^{-1/2},\qquad \mathcal D=z\partial_z.
  \end{aligned}
  \label{eq:pc_Z_closed}
\end{equation}
The same finite-differential generator gives the pair-coherence moment,
\begin{equation}
  M_{\pc}
  =
  \frac{2T\tanh r}{zZ_{\pc}}\mathcal D
  \left[
  \left(T-2R\mathcal D\right)
  \left(T-2R(\mathcal D-1)\right)G(z)
  \right].
  \label{eq:pc_M_differential}
\end{equation}
Although Eq.~\eqref{eq:pc_M_differential} contains an explicit factor $1/z$, its numerator vanishes linearly with $z$. At $r=0$ it is defined by the continuous limit $M_{\pc}=0$; the arbitrary-order form in Appendix~\ref{app:moments} provides the equivalent nonsingular continuation.
The same finite-differential construction therefore supplies $P_j^{(m)}$,
$N_j^{(m)}$, $V_j^{(m)}$, and $M_j^{(m)}$ without an analytic Fock-space
truncation. Appendix~\ref{app:moments} gives the arbitrary-order generators.
These moments define the theory used below. The order-resolved numerical scans
also evaluate the corresponding finite-transmissivity number coefficients in a
high-cutoff Fock representation as an independent implementation; convergence is verified by increasing the Fock cutoff until the plotted quantities are stable.

\subsection{Phase encoding and photon-number resources}

The first OPA is represented by
\begin{equation}
  S_2(g)=\exp\left[g(a^\dagger b^\dagger-ab)\right],
  \label{eq:s2}
\end{equation}
with real gain $g$. We define $C=\cosh g$ and $S=\sinh g$. The Heisenberg transformations are
\begin{align}
  S_2^\dagger(g)aS_2(g)&=Ca+Sb^\dagger,\\
  S_2^\dagger(g)bS_2(g)&=Cb+Sa^\dagger .
  \label{eq:s2_heisenberg}
\end{align}
After the first OPA, a phase shift is applied to mode $a$:
\begin{equation}
  U_\phi=\exp(-\ii\phi \hat n_a).
  \label{eq:phase}
\end{equation}
This sign convention gives
$U_\phi^\dagger aU_\phi=\ee^{-\ii\phi}a$ and
$U_\phi^\dagger a^\dagger U_\phi=\ee^{\ii\phi}a^\dagger$.
The alternative convention $\exp(\ii\phi\hat n_a)$ is equivalent to
$\phi\mapsto-\phi$ and leaves the QFI and phase sensitivity unchanged.
The second OPA is chosen as $S_2(-g)$, so the total ideal interferometer is
\begin{equation}
  U_{\rm SU(1,1)}(\phi)=S_2(-g)U_\phi S_2(g).
\end{equation}
At $\phi=0$ the two OPAs cancel exactly.

The non-Gaussian operation changes the photon statistics entering the
interferometer. A consistent comparison therefore requires the photon numbers
at the phase-encoding plane, immediately after the first OPA and before the
phase shifter. We track both the total two-mode photon number and the
sensing-arm photon number; neither depends on $\phi$. The total photon number is
\begin{equation}
  \bar N_{\rm enc}^{(j,m)}
  =
  \langle S_2^\dagger(g)(\hat n_a+\hat n_b)S_2(g)\rangle_{\psi_{\rm in}^{(j,m)}} .
\end{equation}
Using Eq.~\eqref{eq:s2_heisenberg} and $\langle b\rangle_j^{(m)}=0$, one obtains
\begin{equation}
  \bar N_{\rm enc}^{(j,m)}
  =
  \cosh(2g)\left(|\alpha|^2+N_j^{(m)}\right)
  +2\sinh^2 g.
  \label{eq:Nphi}
\end{equation}
The arm-resolved photon numbers are
\begin{align}
  \bar n_{a,{\rm enc}}^{(j,m)}
  &=
  C^2|\alpha|^2+S^2(N_j^{(m)}+1),\nonumber\\
  \bar n_{b,{\rm enc}}^{(j,m)}
  &=
  C^2N_j^{(m)}+S^2(|\alpha|^2+1),
  \label{eq:arm_photon_numbers}
\end{align}
with $\bar N_{\rm enc}^{(j,m)}=\bar n_{a,{\rm enc}}^{(j,m)}+\bar n_{b,{\rm enc}}^{(j,m)}$.
The total $\bar N_{\rm enc}$ characterizes the correlated two-mode probe,
whereas only mode $a$ passes through the phase shifter and has exposure
$\bar n_{a,{\rm enc}}$. Thus increasing $m$ changes both the ideal information
  and the successful-probe resources before any constrained comparison. Fixing
$\bar N_{\rm enc}$ and fixing $\bar n_{a,{\rm enc}}$ define distinct resource
constraints and must be stated separately.

Equations~\eqref{eq:Nphi} and \eqref{eq:arm_photon_numbers} define the
photon-number resources used in later comparisons. A fixed-preparation comparison keeps
  $|\alpha|$, $r$, $g$, and the module transmissivity unchanged. A conditional-probe comparison must instead state whether it fixes the total two-mode energy
$\bar N_{\rm enc}$ or the sensing-arm exposure $\bar n_{a,{\rm enc}}$. In every case, the
  conditional information is derived first; preparation probability is then included either in the diagnostic difference of Eq.~\eqref{eq:probability_weighted_difference} or in the per-attempt information $P_jF_j$.

\section{Ideal conditional phase estimation with non-Gaussian states}

\subsection{Conditional quantum Fisher information}
\label{sec:qfi}

For a pure state and a unitary phase shift generated by $\hat n_a$, the QFI is \cite{Braunstein1994,Paris2009}
\begin{equation}
  F_Q^{(j,m)}=4\Delta^2_{\Psi_j^{(m)}}\hat n_a,
  \qquad
  \ket{\Psi_j^{(m)}}=S_2(g)\ket{\psi_{\rm in}^{(j,m)}}.
  \label{eq:qfi_prephase}
\end{equation}
Equivalently,
\begin{equation}
  F_Q^{(j,m)}
  =
  4\Delta^2_{\psi_{\rm in}^{(j,m)}}\hat G_a,
  \qquad
  \hat G_a=S_2^\dagger(g)\hat n_aS_2(g).
\end{equation}
From Eq.~\eqref{eq:s2_heisenberg},
\begin{equation}
  \hat G_a
  =
  C^2\hat n_a+S^2(\hat n_b+1)
  +CS(a^\dagger b^\dagger+ab).
  \label{eq:Ga}
\end{equation}
At an arbitrary fixed order $m$, the input is a pure product state, with mode $a$ in $\ket{\alpha}$ and mode $b$ in $\ket{\chi_j^{(m)}}$. Since $\langle b\rangle_j^{(m)}=0$, the mean value is
\begin{equation}
  \langle \hat G_a\rangle
  =
  C^2|\alpha|^2+S^2(N_j+1).
\end{equation}
The variance separates into three nonzero pieces. The coherent state gives $\Delta^2_{\alpha}\hat n_a=|\alpha|^2$. The non-Gaussian squeezed input contributes $V_j$. The two-mode OPA cross term contributes
\begin{align}
  &\left\langle(a^\dagger b^\dagger+ab)^2\right\rangle
  \nonumber\\
  &\quad=
  \alpha^{\ast 2}M_j^\ast+\alpha^2M_j
  +|\alpha|^2N_j+(|\alpha|^2+1)(N_j+1)
  \nonumber\\
  &\quad=
  2\operatorname{Re}(\alpha^2M_j)
  +2|\alpha|^2N_j+|\alpha|^2+N_j+1 .
  \label{eq:X2}
\end{align}
Product structure and $\langle b\rangle_j=0$ imply
$\operatorname{Cov}_{\rm s}(\hat n_a,\hat n_b)
=\operatorname{Cov}_{\rm s}(\hat n_a,\hat X)=0$, where
$\hat X=a^\dagger b^\dagger+ab$ and
$\operatorname{Cov}_{\rm s}(A,B)=
\langle\{\Delta A,\Delta B\}\rangle/2$.
The remaining covariance is
\begin{equation}
  \begin{aligned}
  \mathcal C_j^{(m)}
  &=
  \operatorname{Cov}_{\rm s}(\hat n_b,\hat X)\\
  &=
  \frac{1}{2}
  \left[
  \alpha^*\langle\{\hat n_b,b^\dagger\}\rangle_j^{(m)}
  {}+\alpha\langle\{\hat n_b,b\}\rangle_j^{(m)}
  \right].
  \end{aligned}
  \label{eq:number_weighted_covariance}
\end{equation}
Thus a general \emph{pure} mode-$b$ input satisfying $\langle b\rangle=0$ has
\begin{widetext}
\begin{equation}
  F_{Q,\mathrm{pure}}^{(j,m)}
  =
  4\left[
  C^4|\alpha|^2
  +S^4 V_j^{(m)}
  +C^2S^2
  \left(
  2|\alpha|^2N_j^{(m)}+|\alpha|^2+N_j^{(m)}+1
  +2\operatorname{Re}(\alpha^2M_j^{(m)})
  \right)
  +2CS^3\mathcal C_j^{(m)}
  \right].
  \label{eq:FQ_general_pure}
\end{equation}
\end{widetext}
For the pure parity-eigenstate conditional inputs of this work,
Eq.~\eqref{eq:zero_first_moment} gives $\mathcal C_j^{(m)}=0$. Therefore
\begin{widetext}
\begin{equation}
  F_Q^{(j,m)}
  =
  4\left[
  C^4|\alpha|^2
  +S^4 V_j^{(m)}
  +C^2S^2
  \left(
  2|\alpha|^2N_j^{(m)}+|\alpha|^2+N_j^{(m)}+1
  +2\operatorname{Re}(\alpha^2M_j^{(m)})
  \right)
  \right].
  \label{eq:FQ_main}
\end{equation}
\end{widetext}
Equation~\eqref{eq:FQ_main} is the main analytic result for pure parity-eigenstate inputs at fixed operation order $m$. Its three terms have clear physical origins: amplified coherent number noise (first term), OPA-amplified photon-number fluctuation of the prepared squeezed input (second term), and OPA-created correlation between the coherent input and the pair-coherence moment (third term). Thus $m$ enters the conditional bound through the prepared-state moments, not only through the heralding probability.
Equation~\eqref{eq:FQ_main} is specific to the pure parity-eigenstate conditional inputs used here; lossy mixed probes require the symmetric-logarithmic-derivative QFI.

The pair-coherence contribution is maximized when
\begin{equation}
  \arg M_j^{(m)}+2\arg\alpha=0
  \quad \mathrm{mod}\;2\pi.
  \label{eq:phase_matching}
\end{equation}
All numerical results below use this phase-matching condition. It can be
implemented by choosing the coherent phase relative to the squeezing phase; we discuss the phase-mismatched case only to identify where the interference term enters.
A mismatch reduces $\operatorname{Re}(\alpha^2M_j^{(m)})$ and can remove a
conditional advantage without changing the heralding probability.

For single-photon PS and PA, substitution of Eq.~\eqref{eq:ps_pa_moments}
into Eq.~\eqref{eq:FQ_main} gives the closed finite-transmissivity result
\begin{widetext}
\begin{equation}
  F_Q^{(\ps,1)}=F_Q^{(\pa,1)}
  =4\left\{
  C^4|\alpha|^2
  +\frac{6S^4z}{(1-z)^2}
  +C^2S^2\left[
  \frac{3|\alpha|^2(1+\sqrt z)}{1-\sqrt z}
  +\frac{2+z}{1-z}
  \right]
  \right\},
  \qquad z=T^2\tanh^2r .
  \label{eq:FQ_single_pspa}
\end{equation}
\end{widetext}
Their equality concerns the normalized conditional state; their success
probabilities remain different. For single-photon PC, the corresponding exact
expression is
\begin{equation}
  \begin{aligned}
  F_Q^{(\pc,1)}=4\bigl\{
  &C^4|\alpha|^2+S^4V_{\pc}\\
  &+C^2S^2\bigl[(2|\alpha|^2+1)N_{\pc}
  +|\alpha|^2+1\\
  &\hspace{4.5em}{}+2|\alpha|^2M_{\pc}\bigr]\bigr\},
  \end{aligned}
  \label{eq:FQ_single_pc}
\end{equation}
where $N_{\pc}$, $V_{\pc}$, and $M_{\pc}$ are the finite-differential
functions in Eqs.~\eqref{eq:pc_NV}--\eqref{eq:pc_M_differential}. Equations
\eqref{eq:FQ_single_pspa} and \eqref{eq:FQ_single_pc} are therefore analytic
functions of $T$, rather than results of a separate numerical state model.

The Gaussian benchmark follows by inserting Eq.~\eqref{eq:sv_moments} into
Eq.~\eqref{eq:FQ_main}. With $n_s=\sinh^2r$ and the phase-matching condition,
\begin{equation}
  \begin{aligned}
  F_Q^{(\mathrm G)}=4\bigl[&C^4|\alpha|^2
  +2S^4n_s(n_s+1)\\
  &+C^2S^2\left(|\alpha|^2\ee^{2r}+n_s+1\right)\bigr].
  \end{aligned}
  \label{eq:FQ_gaussian}
\end{equation}
This expression is also obtained directly from
$4\operatorname{Var}[S_2^\dagger(g)\hat n_aS_2(g)]$. Its decomposition into
input number fluctuations and OPA-induced correlations agrees with the general
SU(1,1) QFI construction in Ref.~\cite{Gong2016Intramode}. That reference uses
a phase-sum generator, whereas Eq.~\eqref{eq:FQ_gaussian} refers to the
single-arm phase shift in Eq.~\eqref{eq:phase}; the two printed formulas should
therefore be compared only after matching the phase generator.

The fully transmitting boundary gives two distinct checks. PC is an identity
operation at $T=1$:
\begin{equation}
  \begin{aligned}
  \lim_{T\to1}N_{\pc}&=n_s,
  &\lim_{T\to1}V_{\pc}&=2n_s(n_s+1),\\
  \lim_{T\to1}M_{\pc}&=\sqrt{n_s(n_s+1)},
  &\lim_{T\to1}F_Q^{(\pc,1)}&=F_Q^{(\mathrm G)}.
  \end{aligned}
  \label{eq:pc_gaussian_limit}
\end{equation}
For PS and PA, by contrast, $T=1$ is a zero-probability boundary rather than
an identity operation. Their normalized $T\to1^{-}$ limit is
$S_b(\xi)\ket{1}$, with
\begin{equation}
  \begin{aligned}
  N_{\ps,\pa}&=3n_s+1,\qquad
  V_{\ps,\pa}=6n_s(n_s+1),\\
  M_{\ps,\pa}&=3\sqrt{n_s(n_s+1)}.
  \end{aligned}
  \label{eq:ideal_pspa_moments}
\end{equation}
Substitution gives
\begin{widetext}
\begin{equation}
  \begin{aligned}
  F_{Q,T\to1^-}^{(\ps,\pa;1)}=4\bigl\{
  &C^4|\alpha|^2+6S^4n_s(n_s+1)\\
  &+C^2S^2\bigl[(6n_s+3)|\alpha|^2+3n_s+2
  +6|\alpha|^2\sqrt{n_s(n_s+1)}\bigr]\bigr\}.
  \end{aligned}
  \label{eq:FQ_ideal_pspa_limit}
\end{equation}
\end{widetext}
This is the squeezed-single-photon result implied by
Eq.~\eqref{eq:one_photon_equiv}. It recovers the ideal photon-subtracted and
photon-added input-state limit considered in Refs.~\cite{Gong2016Intramode,Guo2018PhotonAddedSU11},
after the same phase generator is adopted. The conditional QFI remains finite
although $P_{\ps}$ and $P_{\pa}$ vanish as $T\to1^-$. Thus only PC returns to
the Gaussian QFI at full transmission; PS and PA do not.

We compare conditional QFI by the difference
\begin{equation}
  \delta F_Q^{(j,m)}
  =F_Q^{(j,m)}-F_Q^{(\mathrm G)}.
  \label{eq:qfi_difference}
\end{equation}
Thus $\delta F_Q^{(j,m)}>0$ denotes an increase in encoded phase information.

The transmissivity dependence follows from the same formula, without a separate numerical model. For single-photon PS and PA at a common $T$, Eq.~\eqref{eq:ps_pa_moments} depends on $z=T^2\tanh^2r$ and gives
\begin{equation}
 \begin{aligned}
 \frac{dN_{\ps,\pa}}{dz}&=\frac{3}{(1-z)^2},&
 \frac{dV_{\ps,\pa}}{dz}&=\frac{6(1+z)}{(1-z)^3},\\
 \frac{dM_{\ps,\pa}}{dz}&=\frac{3(1+z)}{2\sqrt z(1-z)^2}.&&
 \end{aligned}
 \label{eq:ps_pa_T_derivatives}
\end{equation}
All three derivatives are positive for $0<z<1$. Under the phase-matching condition in Eq.~\eqref{eq:phase_matching}, every moment-dependent coefficient in Eq.~\eqref{eq:FQ_main} is also positive. Since $dz/dT=2T\tanh^2r>0$, the single-photon PS/PA conditional QFI increases monotonically with $T$. The corresponding parity result is obtained after deriving its moment formula in Sec.~III\,B.

PC has no analogous monotonicity. Its diagonal coefficient in Eq.~\eqref{eq:cat_amp} contains interfering powers of $T$ and $1-T$, so its conditional moments can have interior extrema and finite Gaussian-beating intervals.

Unlike analyses that begin with one chosen input family \cite{Li2016ParitySU11,Gao2016Lossy,Xu2023PhotonOps}, Eq.~\eqref{eq:FQ_main} leaves the operation abstract until the three moments are inserted. Its domain is the pure parity-eigenstate family considered here. A more general pure mode-$b$ input with $\langle b\rangle=0$ requires Eq.~\eqref{eq:FQ_general_pure}, and a mixed input requires the symmetric-logarithmic-derivative QFI. The moment formula separates three effects: OPA amplification of $V_j$, phase-sensitive interference through $M_j$, and the resource constraint that fixes the relevant photon budget.

\begin{figure*}[!t]
\includegraphics[width=0.985\textwidth]{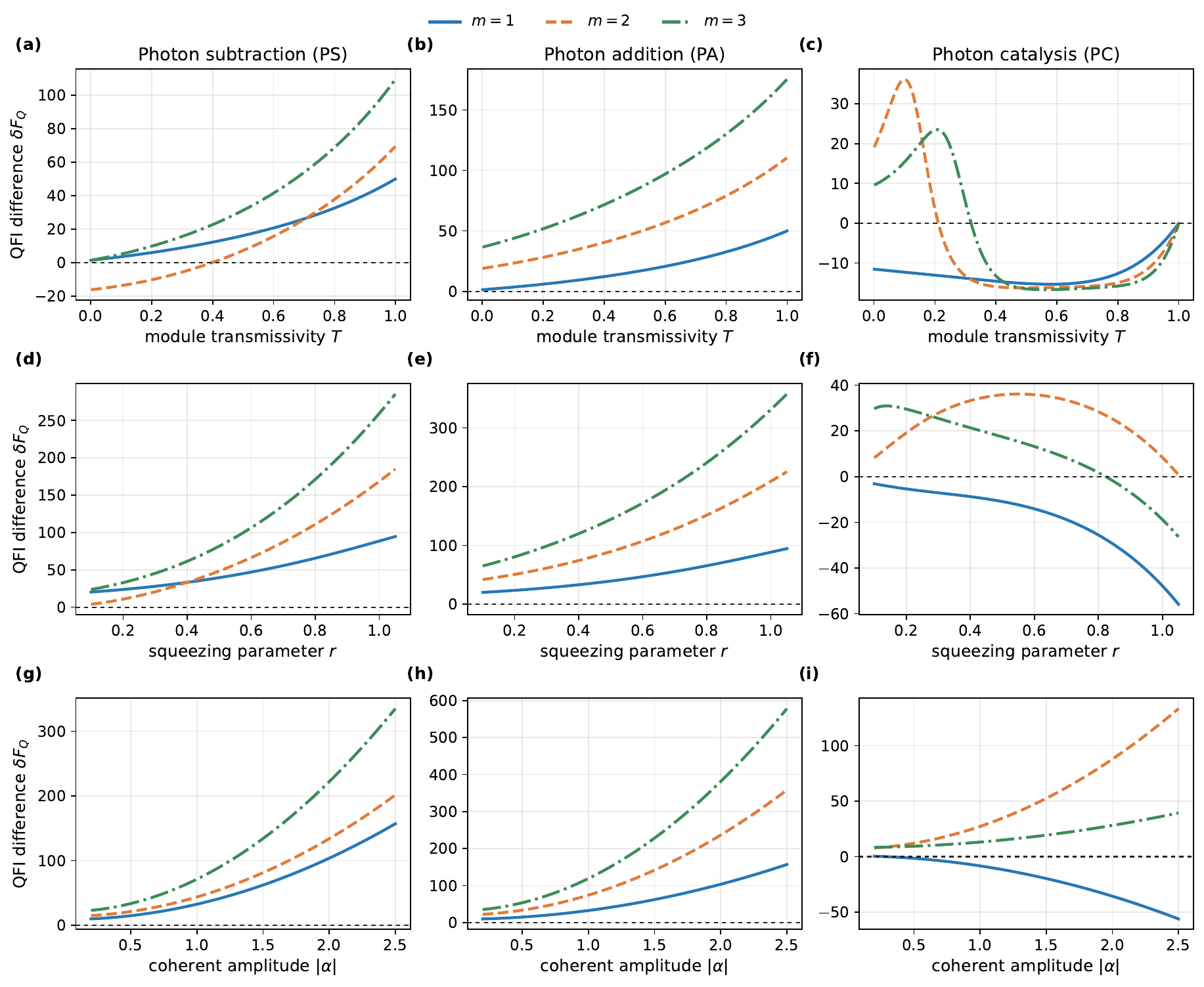}
\caption{Conditional QFI difference $\delta F_Q^{(j,m)}=F_Q^{(j,m)}-F_Q^{(\mathrm G)}$ for PS, PA, and PC at $m=1,2,3$ and $g=0.75$. The first row varies the module transmissivity $T$ at $r=0.55$ and $|\alpha|=1.2$. The second row varies $r$ at $|\alpha|=1.2$; the third row varies $|\alpha|$ at $r=0.55$. The last two rows use $T=0.93$ for PS and PA and $T=0.10$ for PC. Positive values denote a conditional QFI improvement. Color and line style jointly identify the operation order; heralding probability is not included.}
\label{fig:parameter_scans}
\end{figure*}

Figure~\ref{fig:parameter_scans} collects the QFI difference after the
analytic moments have been inserted. Its first row gives the transmissivity
dependence at fixed $r=0.55$ and $|\alpha|=1.2$. For these parameters, PS
improves on the Gaussian QFI throughout $0<T<1$ for $m=1$ and $m=3$, while its
$m=2$ interval is $0.3980<T<1$. PA has a positive QFI difference throughout
$0<T<1$ for $m=1,2,3$. Single-photon PC has no positive interval; its
multi-photon intervals are $0<T<0.2113$ for $m=2$ and $0<T<0.3185$ for $m=3$.
The quoted boundaries are the zeros of the analytic QFI difference for the
stated preparation parameters; they are not universal thresholds.

The second and third rows of Fig.~\ref{fig:parameter_scans} show the dependence
on $r$ and $|\alpha|$, respectively. They use $T=0.93$ for PS and PA, where the
high-transmissivity conditional enhancement is visible, and $T=0.10$ for PC,
inside its low-transmissivity QFI-improvement window. Thus the PC $m=2$ and
$m=3$ curves exhibit a clear positive difference over substantial ranges of
$r$ and $|\alpha|$, whereas the single-photon PC curve remains below the
Gaussian reference.

The three rows distinguish two enhancement mechanisms. For PS and PA,
increasing $T$ weakens the tapping but increases the conditional contributions
of $N_j^{(m)}$, $V_j^{(m)}$, and the phase-matched correlation
$M_j^{(m)}$ in Eq.~\eqref{eq:FQ_main}. This accounts for the strong high-$T$
enhancement, especially for PA and higher operation orders. The PS $m=2$
threshold also shows that a higher order alone does not guarantee an
advantage: the moment changes must be large enough to overcome the Gaussian
reference. PC is different because its Fock coefficients contain competing
powers of $T$ and $1-T$. At $T=0.10$, the $m=2$ and $m=3$ operations act as
selective number filters and give positive QFI differences only over finite
windows in $r$ and $|\alpha|$, whereas the $m=1$ curve remains negative. The
increase with $|\alpha|$ follows from the $|\alpha|^2$ and
$\operatorname{Re}(\alpha^2M_j^{(m)})$ terms, but it does not remove the
operation-dependent sign. Thus PS/PA are best represented by a high-$T$ scan,
while PC requires a low-$T$ scan to reveal its conditional enhancement.

\subsection{Parity-detection-based phase sensitivity}

The measured observable is the parity of one output mode, a standard readout in
interferometric phase estimation \cite{Li2016ParitySU11,Gao2016Lossy},
\begin{equation}
  \Pi_b=(-1)^{\hat n_b}.
  \label{eq:ideal_parity_definition}
\end{equation}
In an ideal balanced SU(1,1) interferometer, the inverse OPA converts small phase changes into an even--odd fringe near the dark point $\phi=0$. Parity is proportional to the single-mode Wigner function at the origin and is therefore sensitive to non-Gaussian interference \cite{Zhang2021NCO}. It is nevertheless a binary, single-mode measurement, so its Fisher information must be compared with the QFI explicitly. In the ideal balanced model, the two OPAs cancel at $\phi=0$ for every input, and the dark fringe remains at $\phi=0$.
The parity signal for operation $j$ at order $m$ is
\begin{equation}
  \langle\Pi_b\rangle_j^{(m)}(\phi)
  =
  \bra{\psi_{\rm in}^{(j,m)}}
  U_{\rm SU(1,1)}^\dagger(\phi)\Pi_b
  U_{\rm SU(1,1)}(\phi)
  \ket{\psi_{\rm in}^{(j,m)}}.
  \label{eq:parity_signal}
\end{equation}
For a binary parity measurement, the probabilities are
\begin{equation}
  p_\pm(\phi)=\frac{1\pm \langle \Pi_b\rangle(\phi)}{2}.
\end{equation}
The corresponding classical Fisher information is
\begin{equation}
  F_C^\Pi(\phi)
  =
  \frac{\left[\partial_\phi\langle\Pi_b\rangle(\phi)\right]^2}
  {1-\langle\Pi_b\rangle^2(\phi)}.
  \label{eq:FC_parity}
\end{equation}
The error-propagation phase sensitivity is
\begin{equation}
  \Delta\phi_\Pi
  =
  \frac{\sqrt{1-\langle\Pi_b\rangle^2}}
  {|\partial_\phi\langle\Pi_b\rangle|}
  =
  \frac{1}{\sqrt{F_C^\Pi}}.
  \label{eq:error_prop}
\end{equation}
At a common operating phase, the conditional improvement over the Gaussian
input is measured by
\begin{equation}
  D_\Pi^{(j,m)}(\phi)
  =\Delta\phi_\Pi^{(\mathrm G)}(\phi)
  -\Delta\phi_\Pi^{(j,m)}(\phi).
  \label{eq:parity_sensitivity_difference}
\end{equation}
Positive $D_\Pi^{(j,m)}$ means that the non-Gaussian conditional state has the
smaller phase uncertainty. All comparisons below use this difference.
Parity detection reaches the quantum Cram\'er--Rao bound only when $F_C^\Pi=F_Q$; this condition is checked below.

Because the input is non-Gaussian, a covariance-matrix calculation is insufficient. Nevertheless, a two-mode Fock expansion is unnecessary for the ideal parity signal. We use $\Pi_b=(2\pi)^{-1}\int d^2\zeta\,D_b(\zeta)$ and the IWOP Gaussian integration method \cite{Fan1992PRA,Fan2008IWOP}. Back-propagating the displacement through the two OPAs gives
$U_{\rm SU(1,1)}^\dagger D_b(\zeta)U_{\rm SU(1,1)}=D_a(A_\phi^{(0)}\zeta^*)D_b(B_\phi^{(0)}\zeta)$. Appendix~\ref{app:ideal_parity_iwop} evaluates the remaining coherent-state trace and Gaussian integral, giving
\begin{align}
  A_\phi^{(0)} &=CS\left(\ee^{\ii\phi}-1\right),
  & B_\phi^{(0)} &=C^2-S^2\ee^{-\ii\phi},\nonumber\\
  \tau_\phi^{(0)}&=A_\phi^{(0)}\alpha^*,
  & \Delta_\phi^{(0)}&=|A_\phi^{(0)}|^2+|B_\phi^{(0)}|^2,
  \label{eq:ideal_parity_kernel_coefficients}
\end{align}
where $A_\phi^{(0)}$ and $B_\phi^{(0)}$ are the backward-propagated
displacements into the coherent and heralded input ports, $\tau_\phi^{(0)}$
collects the coherent-state displacement contribution, and $\Delta_\phi^{(0)}$
sets the Gaussian width.
\begin{equation}
  \langle\Pi_b\rangle_j^{(m)}(\phi)
  =
  \operatorname{Tr}_b\!\left[
  \rho_j^{(m)}\mathcal P_\phi
  \right],
  \label{eq:ideal_parity_effective_signal}
\end{equation}
where $\rho_j^{(m)}=\ket{\chi_j^{(m)}}\bra{\chi_j^{(m)}}$ is the successfully prepared state in mode $b$, and $\mathcal P_\phi={}_a\!\langle\alpha|U_{\rm SU(1,1)}^\dagger(\phi)\Pi_bU_{\rm SU(1,1)}(\phi)|\alpha\rangle_a$ is the pulled-back parity observable after the coherent port has been traced out.
\begin{widetext}
\begin{equation}
  \mathcal P_\phi
  =
  \frac{1}{\Delta_\phi^{(0)}}
  :\exp\!\left[
  \frac{2\bigl(B_\phi^{(0)}b^\dagger-\tau_\phi^{(0)*}\bigr)
  \bigl(\tau_\phi^{(0)}-B_\phi^{(0)*}b\bigr)}
  {\Delta_\phi^{(0)}}
  \right]: .
  \label{eq:ideal_parity_effective_operator}
\end{equation}
\end{widetext}
Equation~\eqref{eq:ideal_parity_effective_signal} is therefore an exact
reduction from a two-mode measurement to a single-mode trace, while
Eq.~\eqref{eq:ideal_parity_effective_operator} is its closed normal-ordered
form. At $\phi=0$, $A_0^{(0)}=0$, $B_0^{(0)}=1$, and $\Delta_0^{(0)}=1$, hence
\begin{equation}
  \mathcal P_0=:\!\exp(-2b^\dagger b)\!:
  =(-1)^{\hat n_b}=\Pi_b .
  \label{eq:ideal_parity_dark_limit}
\end{equation}
Indeed, on $\ket n$ the normally ordered exponential gives
$\sum_{k=0}^{n}(-2)^k\binom nk\ket n=(-1)^n\ket n$. It is therefore exactly
the parity operator defined in Eq.~\eqref{eq:ideal_parity_definition}, as
required when the two OPAs cancel.

The output annihilation operator can also be written explicitly in terms of input operators:
\begin{equation}
  b_{\rm out}(\phi)
  =
  \left(C^2-\ee^{\ii\phi}S^2\right)b
  +CS(1-\ee^{\ii\phi})a^\dagger.
  \label{eq:bout}
\end{equation}
At $\phi=0$, $b_{\rm out}=b$, as required by the inverse-OPA condition. Equation~\eqref{eq:bout} is useful for deriving small-phase expansions and for verifying numerical implementations.

For a pure parity-eigenstate input, the same identity gives
\begin{align}
  \langle\Pi_b\rangle_j^{(m)}(\phi)
  &=
  p_j^{(m)}+\frac{1}{2}\Pi_j^{(m)\prime\prime}(0)\phi^2+O(\phi^3),
  \nonumber\\
  p_j^{(m)}&=\langle\Pi_b\rangle_j^{(m)}(0)=\pm1 .
  \label{eq:parity_dark_expansion}
\end{align}
Although Eq.~\eqref{eq:FC_parity} is then formally $0/0$ at the dark point, substitution of Eq.~\eqref{eq:parity_dark_expansion} gives the finite limit
\begin{equation}
  F_C^{\Pi,j,m}(0)
  \equiv
  \lim_{\phi\rightarrow0}F_C^{\Pi,j,m}(\phi)
  =
  -p_j^{(m)}\Pi_j^{(m)\prime\prime}(0).
  \label{eq:dark_point_cfi}
\end{equation}
Separating the pulled-back generator into its parity-even and parity-odd parts,
$\hat G_+=C^2\hat n_a+S^2(\hat n_b+1)$ and
$\hat G_-=CS\hat X$, gives the explicit moment formula
\begin{equation}
  \begin{aligned}
  F_C^{\Pi,j,m}(0)
  &=
  4C^2S^2
  \big[
  2|\alpha|^2N_j^{(m)}+|\alpha|^2
  \\
  &\qquad
  +N_j^{(m)}+1+2\operatorname{Re}(\alpha^2M_j^{(m)})
  \big].
  \end{aligned}
  \label{eq:dark_point_cfi_moments}
\end{equation}
Consequently,
\begin{equation}
  F_Q^{(j,m)}-F_C^{\Pi,j,m}(0)
  =
  4\left(C^4|\alpha|^2+S^4V_j^{(m)}\right)\geq0 .
  \label{eq:dark_point_gap}
\end{equation}
Parity saturates the QFI at the ideal dark point only if the parity-even generator has zero variance on the input. Equation~\eqref{eq:dark_point_cfi_moments} shows that operation order enters ideal parity extraction through the same conditional moments as the QFI. The ideal numerical calculation evaluates this expression directly. Finite-$\phi$ scans are used to resolve the fringe, and the loss case is treated separately in Sec.~IV.

\begin{figure*}[!t]
\includegraphics[width=0.94\textwidth]{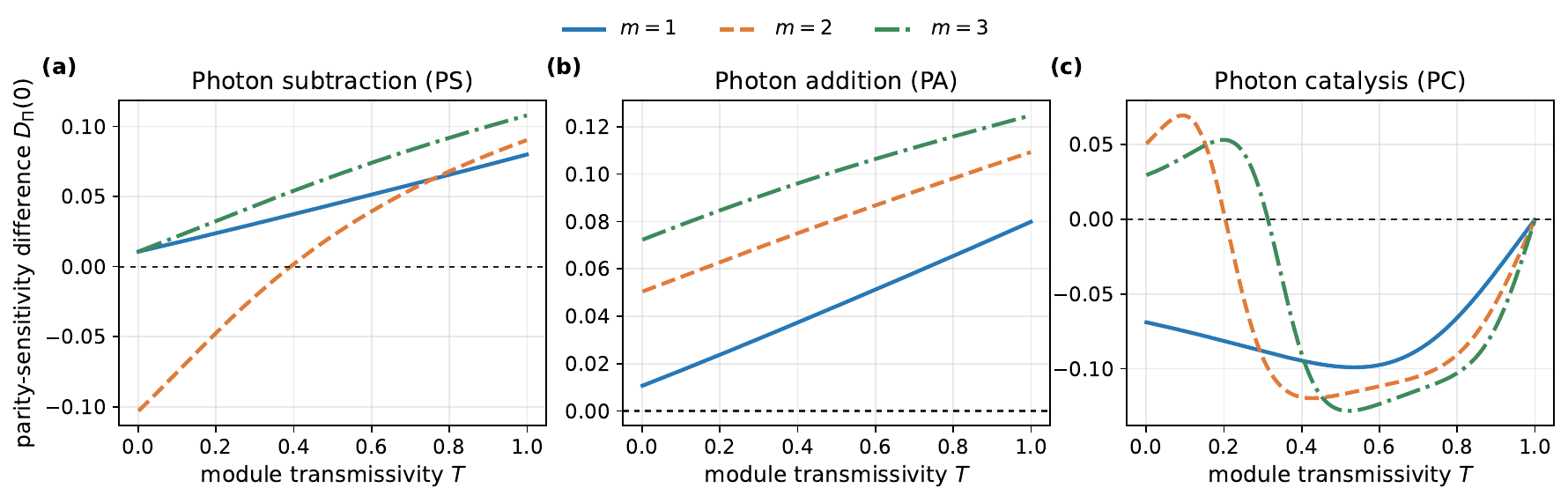}
\caption{Dark-point parity-sensitivity difference $D_\Pi^{(j,m)}(0)=\Delta\phi_\Pi^{(\mathrm G)}(0)-\Delta\phi_\Pi^{(j,m)}(0)$ as a function of the heralding-module transmissivity for PS, PA, and PC at $m=1,2,3$. Parameters are $r=0.55$, $|\alpha|=1.2$, and $g=0.75$. Positive values favor the conditional non-Gaussian probe. Color and line style jointly identify $m=1,2,3$; panels are labeled (a)--(c).}
\label{fig:ideal_T_differences}
\end{figure*}

Table~\ref{tab:ideal_T_windows} collects the transmissivity intervals located from zeros of the moment-based differences and the
peak values underlying Figs.~\ref{fig:parameter_scans} and
\ref{fig:ideal_T_differences}. The table supplements the figures by giving the precise numerical
boundaries.

\begin{table*}[!t]
\caption{Ideal conditional transmissivity intervals for the orders shown in Figs.~\ref{fig:parameter_scans} and \ref{fig:ideal_T_differences}, at fixed $r=0.55$, $|\alpha|=1.2$, and $g=0.75$. The differences follow from the analytic moment formulas and are evaluated with the converged finite-transmissivity Kraus-map implementation; their zeros and interior stationary points are then located numerically. For PS and PA, the entries at the open boundary are limiting values as $T\to1^-$. Success probability is not included.}
\label{tab:ideal_T_windows}
\begin{ruledtabular}
\begin{tabular}{lccccc}
Operation & $m$ & QFI interval & QFI peak $(T,\delta F_Q)$ & parity interval & parity peak $(T,D_\Pi)$\\
\hline
PS/PA & 1 & $0<T<1$ & $T\to1^-: 50.054$ & $0<T<1$ & $T\to1^-: 0.07992$\\
PS & 2 & $0.3980<T<1$ & $T\to1^-: 69.589$ & $0.3920<T<1$ & $T\to1^-: 0.09025$\\
PS & 3 & $0<T<1$ & $T\to1^-: 109.712$ & $0<T<1$ & $T\to1^-: 0.10785$\\
PA & 2 & $0<T<1$ & $T\to1^-: 110.557$ & $0<T<1$ & $T\to1^-: 0.10935$\\
PA & 3 & $0<T<1$ & $T\to1^-: 175.477$ & $0<T<1$ & $T\to1^-: 0.12483$\\
PC & 1 & none & negative, tending to $0$ & none & negative, tending to $0$\\
PC & 2 & $0<T<0.2113$ & $(0.1003,36.159)$ & $0<T<0.2038$ & $(0.0939,0.06956)$\\
PC & 3 & $0<T<0.3185$ & $(0.2069,23.533)$ & $0<T<0.3131$ & $(0.1992,0.05322)$
\end{tabular}
\end{ruledtabular}
\end{table*}

\begin{figure*}[!t]
\includegraphics[width=0.98\textwidth]{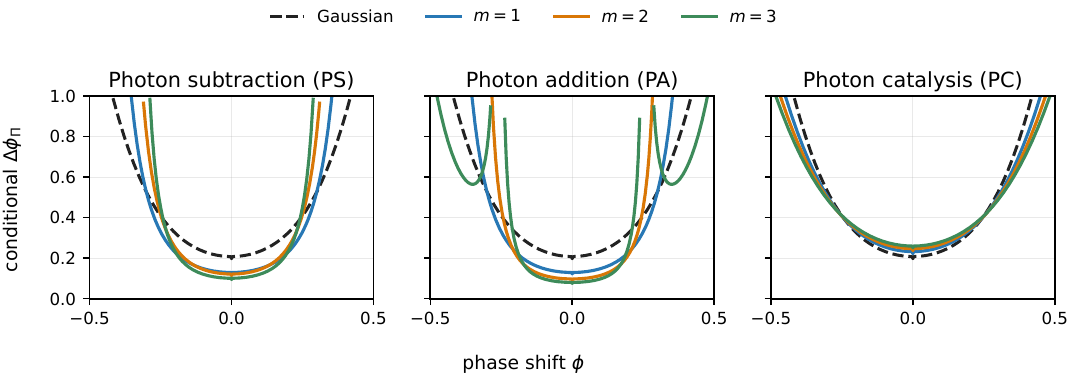}
\caption{Conditional ideal parity phase sensitivity in the local working window, evaluated from the single-mode kernel $\mathcal P_\phi$. All curves are conditioned on a successful heralding event, so the heralding probability is not included. Parameters are $r=0.55$, $|\alpha|=1.2$, $g=0.75$, and $T=0.93$.}
\label{fig:conditional_parity_phase}
\end{figure*}

Figure~\ref{fig:ideal_T_differences} evaluates the dark-point parity-sensitivity
difference from Eq.~\eqref{eq:dark_point_cfi_moments}. For
the parameters stated in the caption, PS has $D_\Pi(0)>0$ throughout $0<T<1$
for $m=1$ and $m=3$, while its $m=2$ interval is $0.3920<T<1$. PA has
$D_\Pi(0)>0$ throughout $0<T<1$ for all three orders. Single-photon PC has no
positive interval; the PC intervals are $0<T<0.2038$ for $m=2$ and
$0<T<0.3131$ for $m=3$. These parity intervals are slightly narrower than the
corresponding QFI intervals in Sec.~III\,A. At $T\to1$, PC approaches the
Gaussian input and $D_\Pi(0)\to0$.

The QFI is phase independent, whereas the parity estimate follows the working curve $\Delta\phi_\Pi(\phi)=\sqrt{1-\langle\hat\Pi_b\rangle_\phi^2}/|\partial_\phi\langle\hat\Pi_b\rangle_\phi|$. Figure~\ref{fig:conditional_parity_phase} resolves this conditional ideal sensitivity for PS, PA, and PC at $m=1,2,3$. It isolates the phase range in which the sensitivity is below one and leaves gaps at stationary parity fringes, where $\partial_\phi\langle\hat\Pi_b\rangle_\phi=0$ and error propagation diverges. These gaps mark poor readout points, not an enhancement. The curves evaluate the single-mode kernel $\mathcal P_\phi$ of Eq.~\eqref{eq:ideal_parity_effective_operator}; representative $d=50,70,90$ convergence values are supplied with the data.

In terms of Eq.~\eqref{eq:parity_sensitivity_difference}, PS and PA have
$D_\Pi^{(j,m)}>0$ in the dark region. The connected interval narrows as the
order increases, apart from a separate $m=3$ PA interval near the outer fringe.
PC has $D_\Pi^{(j,m)}(0)<0$ but becomes positive away from the dark point. At
the displayed parameters, the resolved positive-phase PC intervals begin near
$\phi=0.245$, $0.247$, and $0.253$ for $m=1,2,3$, respectively. PS/PA and PC
therefore provide complementary conditional parity working windows: PS/PA are
favorable near the dark point, whereas PC can be favorable away from the dark point.
The latter is a measurement-specific effect and does not imply a larger QFI or
a better dark-point parity CFI.

\subsection{Difference-based relative performance of non-Gaussian states}

\subsubsection{Probability-weighted sensitivity-difference diagnostic}

The relative-performance measure follows the difference construction of
Ref.~\cite{Kumar2022}. In addition to the parity-sensitivity difference in
Eq.~\eqref{eq:parity_sensitivity_difference}, define the QFI-bound difference
by
\begin{equation}
  D_Q^{(j,m)}
  =\frac{1}{\sqrt{F_Q^{(\mathrm{G})}}}
  -\frac{1}{\sqrt{F_Q^{(j,m)}}}.
  \label{eq:sensitivity_differences}
\end{equation}
These differences are formed before the heralding probability is introduced.
For comparison with the difference construction in Ref.~\cite{Kumar2022}, we also define the probability-weighted diagnostic
\begin{equation}
  \widetilde D_\mu^{(j,m)}
  =P_j^{(m)}D_\mu^{(j,m)},
  \qquad \mu\in\{Q,\Pi\}.
  \label{eq:probability_weighted_difference}
\end{equation}
Here $\mu=Q$ refers to the QFI-bound difference in
Eq.~\eqref{eq:sensitivity_differences}, whereas $\mu=\Pi$ refers to the
parity-sensitivity difference in Eq.~\eqref{eq:parity_sensitivity_difference}.
Because $P_j^{(m)}>0$ for every physical heralding event, $\widetilde D_\mu^{(j,m)}$ has exactly the same sign and the same zero crossings as $D_\mu^{(j,m)}$. It therefore measures the probability-rescaled magnitude and can change the ordering of operations, but it cannot determine whether a conditional improvement survives a per-attempt resource accounting.

\begin{figure*}[!t]
\includegraphics[width=\textwidth]{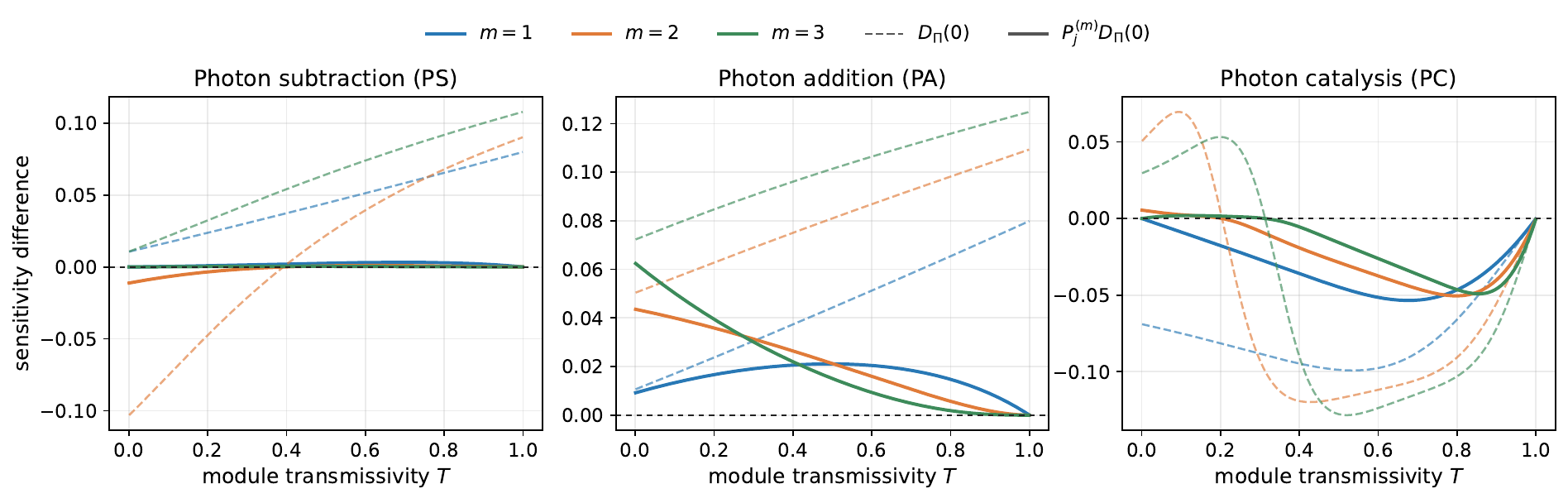}
\caption{Conditional dark-point sensitivity difference $D_\Pi^{(j,m)}(0)$ (dashed curves) and its probability-weighted diagnostic $\widetilde D_\Pi^{(j,m)}(0)=P_j^{(m)}D_\Pi^{(j,m)}(0)$ (solid curves) versus module transmissivity. Parameters are $r=0.55$, $|\alpha|=1.2$, and $g=0.75$. Weighting changes the magnitude, not the sign or zero crossings.}
\label{fig:probability_weighted_differences}
\end{figure*}

Figure~\ref{fig:probability_weighted_differences} shows that probability weighting strongly suppresses the magnitude of the PS and PA differences near $T=1$, where their conditional limits are finite but their success probabilities vanish. It also changes their relative magnitudes: PA gives the largest positive weighted diagnostic among the displayed operations. The positive intervals themselves are unchanged from Fig.~\ref{fig:ideal_T_differences}: single-photon PC has none, whereas $m=2$ and $m=3$ PC retain the same low-$T$ intervals as their conditional differences.

The diagnostic in Eq.~\eqref{eq:probability_weighted_difference} weights an already formed sensitivity difference. By contrast, $P_jF_C^\Pi$ or $P_jF_Q$ is the information per ideal module attempt if a failed herald is assigned zero phase information; the associated effective bound is $[P_jF_j]^{-1/2}$, not $P_j/\sqrt{F_j}$. These quantities must be interpreted according to the experimental timing. With offline heralding and storage, $P_j$ primarily lowers the accepted-probe rate while the conditional information describes each stored probe. With an inline module and equal clock attempts, $P_jF_j$ is the appropriate idealized information per attempt. Neither convention includes ancillary-state generation, detector imperfections, or timing overhead \cite{Combes2014Postselection}. The following constrained benchmarks therefore report conditional and per-attempt information separately.

\subsection{Conditional-probe resource constraints and ideal benchmarks}

We evaluate the derived quantities first at fixed preparation parameters and then under three resource contracts. Protocol I is the fixed-seed module-insertion comparison: it holds $|\alpha|$, $r$, $g$, and the module transmissivity fixed while the squeezed input is replaced by its photon-subtracted, photon-added, or photon-catalyzed version. It isolates the effect of adding a heralding module to a specified preparation chain, but it fixes neither the total two-mode energy nor the sensing-arm photon exposure.

Throughout the three protocols, the required ancilla state and ideal PNR herald
detection are assumed available. PS uses a vacuum ancilla, whereas PA and PC use
an $m$-photon Fock ancilla. The source energy, generation probability, detector
inefficiency, and false heralds associated with these ancillas are not included
in the photon-number resources. This convention follows conditional-state
comparisons in non-Gaussian interferometry \cite{Kumar2022,Combes2014Postselection}.
The probability-weighted diagnostics below follow the same resource caution for postselected metrology.

Protocol II fixes the total two-mode energy of the \emph{successful conditional probe} before phase encoding. For every operation $j$, the coherent amplitude is adjusted so that
\begin{equation}
  \bar N_{\rm enc}^{(j)}
  =
  \bar N_{\rm enc}^{(\mathrm{G})},
  \label{eq:fixed_Nphi_condition}
\end{equation}
where $\bar N_{\rm enc}$ is defined in Eq.~\eqref{eq:Nphi}. This protocol matches the energy of the correlated two-mode probe conditional on heralding, but it does not match the photon number in the phase arm, the mean resource consumed before failed heralds, or the ancillary Fock-state resource. Protocol III instead fixes the sensing-arm photon exposure,
\begin{equation}
  \bar n_{a,{\rm enc}}^{(j)}
  =
  \bar n_{a,{\rm enc}}^{\rm target},
  \qquad
  |\alpha_j|^2
  =
  \frac{\bar n_{a,{\rm enc}}^{\rm target}-S^2(N_j+1)}{C^2},
  \label{eq:fixed_sensing_dose}
\end{equation}
whenever the right-hand side is nonnegative. This is the direct exposure-matched comparison for the single-arm phase object in Eq.~\eqref{eq:phase}; the idler-arm energy and hence $\bar N_{\rm enc}$ may differ.

A fixed-resource slice is not an optimized benchmark. For a resource $R$ and an allowed parameter domain $\mathcal D_j$, we define the fixed-gain conditional optimum
\begin{equation}
  F_{X,j}^{\star}(R)
  =
  \max_{\bm\theta_j\in\mathcal D_j,\,
  R_j(\bm\theta_j)=R}
  F_{X,j}(\bm\theta_j),
  \qquad X\in\{Q,C^\Pi\}.
  \label{eq:optimized_frontier}
\end{equation}
Here the OPA gain $g$ is treated as a common device setting, while the coherent--squeezed allocation is optimized. The value $g=0.75$ used below lies in the moderate-gain regime where the Gaussian conditional QFI is near its unconstrained optimum for the resource levels considered and where the two-mode squeezing is strong enough to produce visible interference fringes without entering the highly depleted pump regime; it is shared by all input families and is not an additional optimization variable. A non-Gaussian advantage under these constraints therefore means an advantage within this fixed-gain coherent-plus-squeezed Gaussian reference family, not over all Gaussian probes or all gains. Protocol I diagnoses insertion of the heralding module, Protocol II fixes the total energy of the successful conditional probe, Protocol III fixes its sensing-arm exposure, and the fixed-gain optimized value gives the strictest conditional-probe benchmark used here. None is a complete laboratory-resource match because failed preparations and ancillary-state generation are not assigned an energy cost.

The preparation, QFI, and parity formulas retain the order index $m$. The conditional-probe benchmarks below use $m=1$ as a representative constrained comparison after the order-resolved screen.

\subsubsection{Operation-order screening and single-photon scope}
\label{sec:order_screen}

The order-resolved scans use $T=0.93$, $r=0.55$, $g=0.75$, and $|\alpha|=1.2$. They vary $m=1,2,3,4$ without matching the conditional energy, so they diagnose operation order rather than a resource advantage. The QFI and dark-point parity values are calculated from Eqs.~\eqref{eq:FQ_main} and \eqref{eq:dark_point_cfi_moments}; finite-phase curves use the exact single-mode kernel $\mathcal P_\phi$.

\begin{figure*}[!t]
\includegraphics[width=0.98\textwidth]{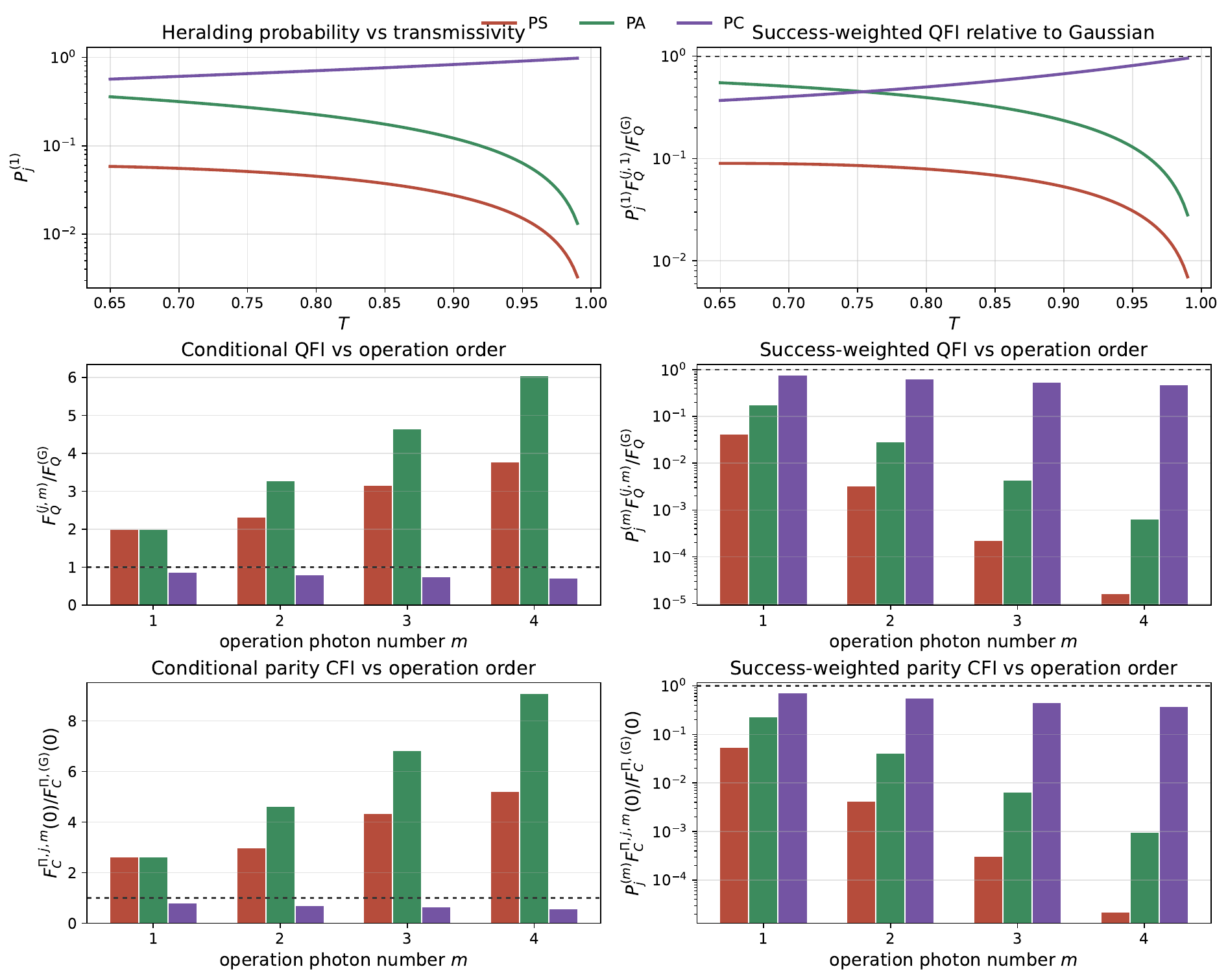}
\caption{Ideal operation-order screen at fixed seed and gain. Top row: $P_j^{(1)}$ and $P_j^{(1)}F_Q^{(j,1)}/F_Q^{(\mathrm G)}$ versus the heralding-module transmissivity $T$. Middle row: conditional and success-weighted QFI versus $m=1,2,3,4$ at $T=0.93$. Bottom row: the corresponding conditional and success-weighted dark-point parity CFI. The middle and bottom rows are normalized by their Gaussian values. These ratios are order-screening diagnostics; the improvement scans in Figs.~\ref{fig:parameter_scans} and \ref{fig:ideal_T_differences} use differences. The dotted line denotes equality with the Gaussian benchmark. The high-$T$ PC curves approach unity from below because $K_{\pc}^{(m)}\rightarrow I$.}
\label{fig:operation_scans}
\end{figure*}

Figure~\ref{fig:operation_scans} collects the operation-order screen. PS and PA increase the conditional QFI and dark-point parity CFI at the displayed fixed seed, generally more strongly at higher $m$. Their success-weighted information instead decreases rapidly because $P_{\ps,\pa}^{(m)}\propto(1-T)^m$ near $T=1$. PC has $P_{\pc}^{(m)}\rightarrow1$, but its conditional and weighted dark-point values remain below the Gaussian benchmark throughout this screen.

At $T=0.93$, the per-attempt quantities $P_jF_Q$ and $P_jF_C^\Pi(0)$ of PS and PA fall below their Gaussian values even though the corresponding conditional quantities are larger. A finite-phase comparison based on $[P_jF_C^\Pi(\phi)]^{-1/2}$ likewise requires a specified clock-attempt convention and working phase; it is not inferred from $P_jD_\Pi$. These facts, together with the order-resolved screen, motivate the representative $m=1$ conditional-probe benchmarks below; the internal-loss comparison in Sec.~IV separately displays $m=1,2,3$.

\subsubsection{Protocol I: fixed-seed module insertion}

Table~\ref{tab:mzi_standard} follows the module-insertion convention of Ref.~\cite{Kumar2022} after the operation-order screen has selected $m=1$. The parameters $|\alpha|=1.2$, $r=0.55$, $g=0.75$, and $T=0.93$ are fixed for all input states. The squeezing phase is chosen so that $\operatorname{Re}(\alpha^2M_j)>0$.

\begin{table*}[t]
\caption{Protocol I: fixed-seed module insertion and fixed SU(1,1) gain. Neither the total photon number $\bar N_{\rm enc}$ nor the sensing-arm photon exposure $\bar n_{a,{\rm enc}}$ is fixed across rows.}
\label{tab:mzi_standard}
\begin{ruledtabular}
\begin{tabular}{lcccccc}
Input & $P_j$ & $\bar N_{\rm enc}$ & $F_Q$ & $P_jF_Q$ & $F_C^\Pi(0)$ & $P_jF_C^\Pi(0)$ \\
\hline
Gaussian SV & $1$ & $5.53$ & $43.48$ & $43.48$ & $25.66$ & $25.66$\\
PS & $2.04\times10^{-2}$ & $9.04$ & $86.59$ & $1.76$ & $66.53$ & $1.35$\\
PA & $8.74\times10^{-2}$ & $9.04$ & $86.59$ & $7.57$ & $66.53$ & $5.82$\\
PC & $8.76\times10^{-1}$ & $5.17$ & $37.25$ & $32.62$ & $20.30$ & $17.78$
\end{tabular}
\end{ruledtabular}
\end{table*}

Under this protocol, PS and PA increase both conditional quantities while raising the total energy before phase encoding from about $5.53$ to $9.04$ and the sensing-arm photon exposure from $3.316$ to $4.327$. Success weighting then reduces their information per ideal module attempt. PC does not improve the conditional quantities at this point.

\subsubsection{Protocol II: fixed total two-mode energy on a fixed-seed slice}

Table~\ref{tab:fixed_resource} repeats the comparison with total $\bar N_{\rm enc}=9$ fixed. For each input operation, $|\alpha|$ is adjusted using Eq.~\eqref{eq:Nphi}, while $r=0.55$, $g=0.75$, and $T=0.93$ are unchanged. The resulting rows form a fixed-$r$ slice through the total-energy constraint.

\begin{table*}[t]
\caption{Protocol II: fixed total two-mode energy of the successful conditional probe before phase encoding, on the $r=0.55$ preparation slice. Parameters are $m=1$, $g=0.75$, $T=0.93$, and $\bar N_{\rm enc}=9$. The sensing-arm photon exposure, resources consumed by failed heralds, and ancillary Fock-state resource are not fixed; this table is not a fully optimized benchmark.}
\label{tab:fixed_resource}
\begin{ruledtabular}
\begin{tabular}{lcccccc}
Input & $|\alpha|$ & $P_j$ & $F_Q$ & $P_jF_Q$ & $F_C^\Pi(0)$ & $P_jF_C^\Pi(0)$ \\
\hline
Gaussian SV & $1.708$ & $1$ & $80.19$ & $80.19$ & $45.78$ & $45.78$\\
PS & $1.192$ & $2.04\times10^{-2}$ & $85.67$ & $1.74$ & $65.83$ & $1.34$\\
PA & $1.192$ & $8.74\times10^{-2}$ & $85.67$ & $7.49$ & $65.83$ & $5.75$\\
PC & $1.752$ & $8.76\times10^{-1}$ & $72.47$ & $63.47$ & $37.20$ & $32.58$
\end{tabular}
\end{ruledtabular}
\end{table*}

The same fixed-resource rows have the following arm-resolved photon-number distribution after the first OPA:
\begin{table}[t]
\caption{Arm-resolved photon numbers for the fixed-total slice in Table~\ref{tab:fixed_resource}. The two columns add to $\bar N_{\rm enc}=9$ up to rounding, but the sensing-arm photon exposures are unequal.}
\label{tab:fixed_resource_arms}
\begin{ruledtabular}
\begin{tabular}{lcc}
Input & $\bar n_{a,{\rm enc}}$ & $\bar n_{b,{\rm enc}}$\\
\hline
Gaussian SV & $5.791$ & $3.209$\\
PS & $4.296$ & $4.704$\\
PA & $4.296$ & $4.704$\\
PC & $5.944$ & $3.056$
\end{tabular}
\end{ruledtabular}
\end{table}

\begin{figure*}[!t]
\includegraphics[width=0.92\textwidth]{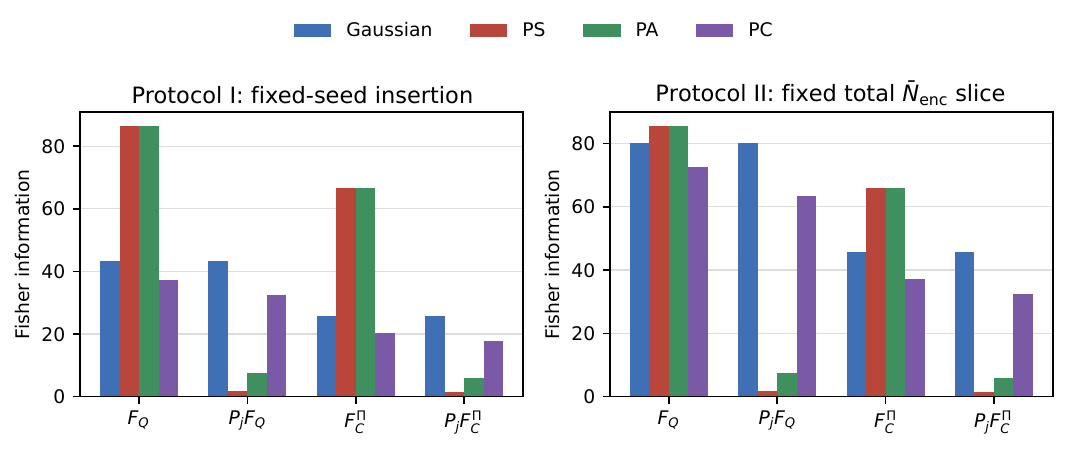}
\caption{Fixed-seed comparisons evaluated from the moments used in Tables~\ref{tab:mzi_standard} and \ref{tab:fixed_resource}. Protocol I fixes the preparation chain; Protocol II fixes the successful conditional-probe energy at $\bar N_{\rm enc}=9$ while retaining $r=0.55$. The parity bars are analytic dark-point values. The success-weighted quantities are the columns labeled $P_jF_Q$ and $P_jF_C^\Pi(0)$; neither panel is an optimized benchmark.}
\label{fig:protocols}
\end{figure*}

Figure~\ref{fig:protocols} summarizes Protocol I and the fixed-total preparation slice; the fixed-gain optima are presented next.

\subsubsection{Metric-specific fixed-gain optima at fixed total energy}

The apparent PS/PA improvement in Table~\ref{tab:fixed_resource} is relative only to the unoptimized Gaussian point at $r=0.55$. We therefore maximize each metric independently over $0\le r\le1.25$ and, for PS/PA/PC, $0.65\le T\le0.995$, while holding $g=0.75$, $m=1$, and $\bar N_{\rm enc}=9$ fixed. For PS the feasible set excludes $r=0$, where the success probability vanishes and the normalized conditional state is undefined. Table~\ref{tab:optimized_total} reports the resulting fixed-gain optima.

\begin{table*}[t]
\caption{Metric-specific fixed-gain optima at fixed total two-mode energy of the successful conditional probe before phase encoding, $\bar N_{\rm enc}=9$. Each column is optimized independently over $m=1$, fixed $g=0.75$, $0\leq r\leq1.25$, and $0.65\leq T\leq0.995$. The PC optima occur at the upper boundary $T=0.995$, where catalysis approaches the identity operation; this restricted weak-tapping domain does not test the low-$T$ filtering windows of Fig.~\ref{fig:ideal_T_differences}.}
\label{tab:optimized_total}
\begin{ruledtabular}
\begin{tabular}{lcccc}
Input & $F_Q^\star$ & $(P_jF_Q)^\star$ & $F_C^{\Pi,\star}(0)$ & $(P_jF_C^\Pi)^\star(0)$\\
\hline
Gaussian SV & $107.569$ & $107.569$ & $73.851$ & $73.851$\\
PS & $86.472$ & $11.544$ & $65.922$ & $8.938$\\
PA & $86.472$ & $30.940$ & $65.922$ & $23.448$\\
PC & $107.546$ & $104.290$ & $73.851$ & $71.740$
\end{tabular}
\end{ruledtabular}
\end{table*}

The fixed-gain Gaussian allocation for conditional QFI is $r=1.06907$ and $|\alpha|=1.26514$, giving $F_Q^{(\mathrm G,\star)}=107.569$. The optimized single-photon PS/PA conditional value is $86.472$, or $0.804$ of this fixed-gain Gaussian value. Thus the approximately $7\%$ increase over the fixed-$r$ Gaussian row in Table~\ref{tab:fixed_resource} is not a Gaussian-beating fixed-total-energy advantage. Within the restricted $m=1$, $r\leq1.25$, $T\geq0.65$ domain, the PC conditional value approaches the fixed-gain Gaussian value only at $T=0.995$, where its Kraus operator is close to the identity. The success-weighted columns give the same conclusion. This common weak-squeezing domain permits a direct PS/PA/PC comparison, but it does not cover all PC parameters.

To determine whether the initial squeezing bound hides a distinct PC branch, we extend the PC scan to $0\leq r\leq2.4$ while retaining $m=1$, $g=0.75$, $0.65\leq T\leq0.995$, and $\bar N_{\rm enc}=9$. The largest conditional QFI found in this domain occurs at $r=1.91428$, $T=0.94669$, and $|\alpha|\simeq0$, where $F_Q=189.890$. At this point the number-variance contribution $4S^4V_{\pc}=170.616$ accounts for $89.85\%$ of the QFI, while the parity-changing term contributes $19.273$. The allocation change near $T\simeq0.974$ separates this high-variance branch from a near-Gaussian branch. This is a conditional branch with high local QFI, not an average-preparation-cost advantage: the pre-herald squeezed input contains about $11.0$ photons, and with $P_{\pc}\simeq0.461$ its seed consumption is about $23.9$ photons per accepted probe before ancillary-photon costs are included.

\begin{figure*}[!t]
\includegraphics[width=0.92\textwidth]{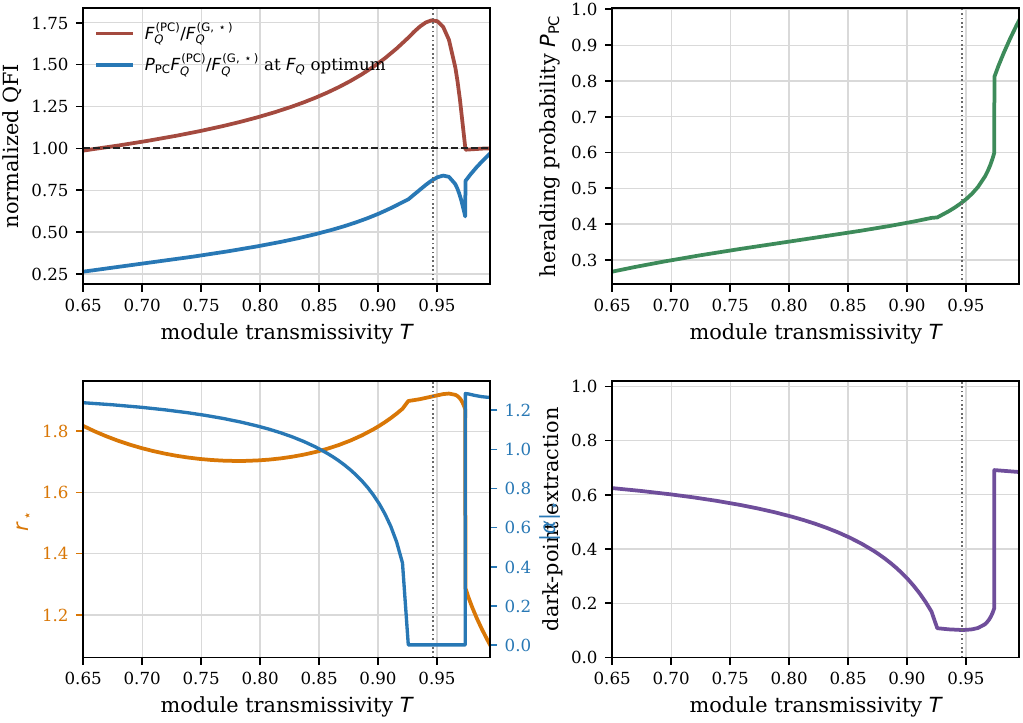}
\caption{Expanded photon-catalysis scan at fixed successful conditional-probe energy $\bar N_{\rm enc}=9$, $g=0.75$, and $m=1$. For every $T$, the PC allocation maximizes conditional $F_Q$ over $0\leq r\leq2.4$, while the energy constraint fixes the coherent amplitude. Panel (a) compares the conditional QFI and its success-weighted value with the fixed-gain Gaussian conditional value. The blue curve is evaluated at the conditional-QFI optimum, not at a separately optimized success-weighted point. Panels (b)--(d) give the heralding probability, optimizing allocation, and dark-point parity extraction. The vertical dotted line marks the largest conditional QFI found in the scanned domain. The branch is a conditional local quantum Fisher information result; average seed and ancillary preparation costs are not fixed.}
\label{fig:extended_pc_frontier}
\end{figure*}

Figure~\ref{fig:extended_pc_frontier} gives the expanded conditional-QFI frontier; its success-weighted QFI remains below the fixed-gain Gaussian value. Table~\ref{tab:extended_pc_frontier} adds the metric-specific success-weighted optima from $r\leq1.25$. Both occur at $T=0.995$, where PC is close to the identity map.

\begin{table*}[t]
\caption{PC benchmarks at $\bar N_{\rm enc}=9$, $g=0.75$, $m=1$, and $0.65\leq T\leq0.995$. The $F_Q$ row uses $0\leq r\leq2.4$; the success-weighted rows use $0\leq r\leq1.25$. All rows use the same fixed gain and conditional-probe energy.}
\label{tab:extended_pc_frontier}
\begin{ruledtabular}
\begin{tabular}{lcccccc}
Objective & $r$ & $T$ & $|\alpha|$ & $P_{\pc}$ & PC optimum & Gaussian value\\
\hline
$F_Q$ & $1.914$ & $0.94669$ & $0$ & $0.461$ & $189.890$ & $107.569$\\
$P_{\pc}F_Q$ & $1.087$ & $0.99500$ & $1.289$ & $0.970$ & $104.290$ & $107.569$\\
$P_{\pc}F_C^\Pi(0)$ & $1.062$ & $0.99500$ & $1.324$ & $0.972$ & $71.740$ & $73.851$
\end{tabular}
\end{ruledtabular}
\end{table*}

At the conditional-QFI optimum, $P_{\pc}F_Q=87.493$ and
$F_C^\Pi(0)/F_Q=0.1015$. This point is useful for diagnosing where the encoded
QFI is stored, but it is not used as a noisy-parity optimum. Internal-loss
parity comparisons are reported separately below with the same measurement
model for all inputs. Since a fixed mean photon number does not constrain the photon-number variance or the distribution tail, this branch does not by itself establish a finite-sample or prior-averaged precision gain.

Two dimensionless diagnostics summarize the ideal resource comparison. The fraction of QFI extracted by dark-point parity is
\begin{equation}
  \mathcal R_\Pi^{(j)}
  =
  \frac{F_C^{\Pi,j}(0)}{F_Q^{(j)}},
  \label{eq:parity_extraction_ratio}
\end{equation}
whereas a success-weighted fixed-total-energy QFI beats the fixed-gain Gaussian value only if
\begin{equation}
  \mathcal A_j^{(Q)}
  =
  \frac{P_jF_Q^{(j)}}{F_Q^{(\mathrm G,\star)}}>1.
  \label{eq:success_threshold}
\end{equation}
The denominator in Eq.~\eqref{eq:success_threshold} is the metric-matched Gaussian optimum at the same $\bar N_{\rm enc}$ and $g$.

\begin{figure*}[!t]
\includegraphics[width=0.92\textwidth]{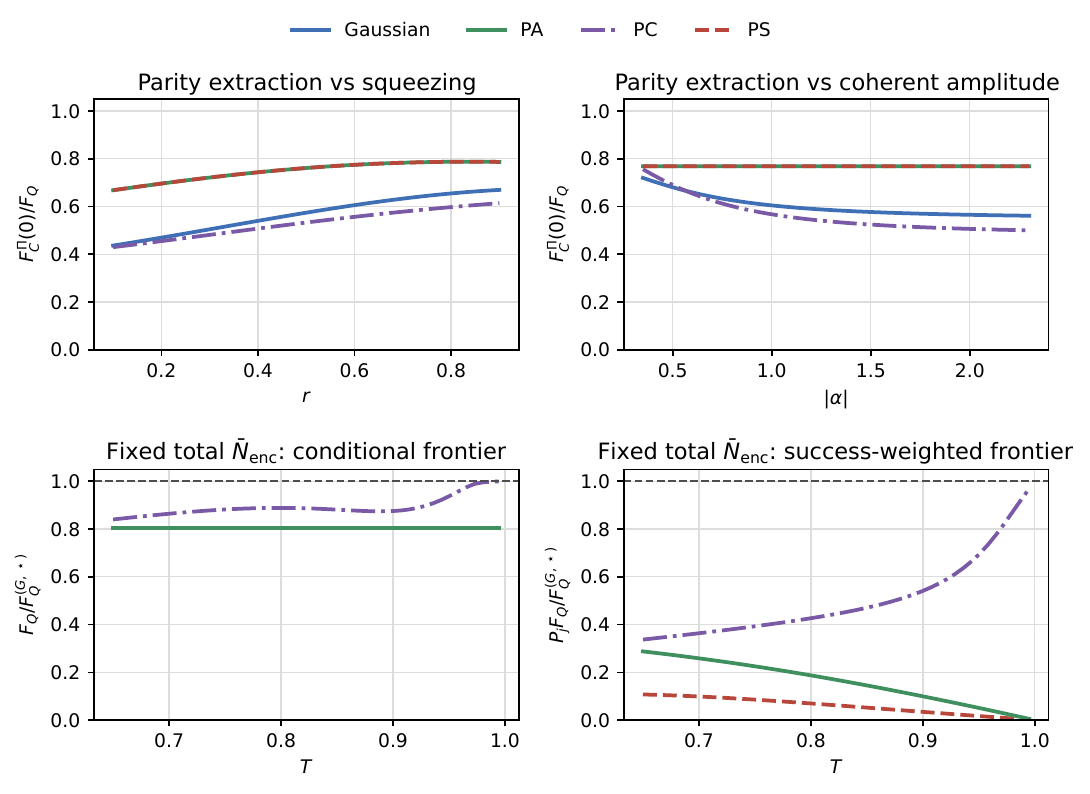}
\caption{Ideal parity extraction and restricted fixed-gain optima at fixed conditional-probe energy. Upper panels: $\mathcal R_\Pi=F_C^\Pi(0)/F_Q$ from the closed moment formulas. Lower panels: at $\bar N_{\rm enc}=9$, fixed $g=0.75$, $m=1$, $0\leq r\leq1.25$, and $0.65\leq T\leq0.995$, $r$ is optimized separately for each $T$ and metric; the denominator is the fixed-gain Gaussian QFI $F_Q^{(\mathrm G,\star)}=107.569$. The dashed line marks equality with that restricted value. The separate expanded-domain PC check is given in Fig.~\ref{fig:extended_pc_frontier}.}
\label{fig:efficiency_fixed_resource}
\end{figure*}

Figure~\ref{fig:efficiency_fixed_resource} shows that parity extracts only the parity-changing part of the generator and does not saturate the QFI in the plotted ranges. In the lower panels, $r\leq1.25$ is optimized separately at each $T$ for the conditional and success-weighted quantities. The optimized PS/PA conditional ratio is approximately $0.804$ throughout the equivalent-state manifold. Restricted-domain PC approaches unity only at the $T=0.995$ identity boundary, and every success-weighted curve remains below the fixed-gain Gaussian value. At $r=0.55$ and $|\alpha|=0.35$, the analytic extraction ratios are $0.768$ for PS/PA and $0.756$ for PC. Equation~\eqref{eq:dark_point_gap} explains why neither reaches unity: coherent number noise and the $b$-mode number variance belong to the parity-even generator sector. Because PC can have a favorable finite-phase parity window away from $\phi=0$, the dark-point parity comparison is a common readout benchmark rather than a phase-optimized PC benchmark.

\subsubsection{Protocol III: fixed sensing-arm photon exposure}

To match the photons incident on the single-arm phase object, Table~\ref{tab:fixed_dose} fixes $\bar n_{a,{\rm enc}}=5.791$, the Gaussian sensing-arm photon exposure in Table~\ref{tab:fixed_resource}, using Eq.~\eqref{eq:fixed_sensing_dose}. The seed parameters remain $r=0.55$, $g=0.75$, and $T=0.93$.

\begin{table*}[t]
\caption{Protocol III: fixed sensing-arm photon exposure $\bar n_{a,{\rm enc}}=5.791$ on the $r=0.55$ preparation slice. The total two-mode energy $\bar N_{\rm enc}$ is allowed to differ and is reported explicitly.}
\label{tab:fixed_dose}
\begin{ruledtabular}
\begin{tabular}{lccccccc}
Input & $|\alpha|$ & $P_j$ & $\bar N_{\rm enc}$ & $F_Q$ & $P_jF_Q$ & $F_C^\Pi(0)$ & $P_jF_C^\Pi(0)$\\
\hline
Gaussian SV & $1.708$ & $1$ & $9.00$ & $80.19$ & $80.19$ & $45.78$ & $45.78$\\
PS & $1.521$ & $2.04\times10^{-2}$ & $11.10$ & $128.97$ & $2.63$ & $99.10$ & $2.02$\\
PA & $1.521$ & $8.74\times10^{-2}$ & $11.10$ & $128.97$ & $11.27$ & $99.10$ & $8.66$\\
PC & $1.726$ & $8.76\times10^{-1}$ & $8.79$ & $70.50$ & $61.74$ & $36.26$ & $31.76$
\end{tabular}
\end{ruledtabular}
\end{table*}

At equal sensing-arm photon exposure, PS/PA again have larger conditional values on this fixed-seed slice, but they use more idler-arm energy and hence a larger total $\bar N_{\rm enc}$. A full fixed-gain optimum at fixed sensing-arm photon exposure therefore requires an additional bound on the idler energy or total preparation energy. Without that second constraint, the fixed-exposure and fixed-total-energy comparisons answer different questions and cannot be ranked by one measure of advantage.
\section{Phase sensitivity under internal loss}

\subsection{Effective parity observable under internal loss}

Internal loss in the phase-sensing arms is modeled by fictitious beam splitters \cite{Marino2012,Gao2016Lossy}:
\begin{align}
  a&\rightarrow \sqrt{\eta_a}\,a+\sqrt{1-\eta_a}\,v_a,\\
  b&\rightarrow \sqrt{\eta_b}\,b+\sqrt{1-\eta_b}\,v_b,
\end{align}
where $v_a$ and $v_b$ are vacuum environment modes. We take the output parity
measurement to be ideal throughout this section; the only noisy
processes are the two internal pure-loss channels inserted before the second
OPA. The pure-loss channel on one mode is
\begin{equation*}
  \mathcal L_\eta(\rho)
  =
  \sum_{\ell=0}^{\infty}L_\ell^{(\eta)}\rho L_\ell^{(\eta)\dagger},
  \qquad
  L_\ell^{(\eta)}
  =
  \sqrt{\frac{(1-\eta)^\ell}{\ell!}}\,
  \eta^{\hat n/2}a^\ell .
\end{equation*}
The complete noisy map can first be specified in channel form as the starting point
\begin{widetext}
\begin{equation*}
  \langle\Pi_b\rangle_{\eta_a,\eta_b}^{(j,m)}(\phi)
  =
  \operatorname{Tr}\left[
  \Pi_b
  S_2(-g)
  \left(\mathcal L_{\eta_a}\otimes\mathcal L_{\eta_b}\right)
  \left(
  U_\phi S_2(g)\rho_{\rm in}^{(j,m)}S_2^\dagger(g)U_\phi^\dagger
  \right)
  S_2^\dagger(-g)
  \right].
\end{equation*}
\end{widetext}
This expression fixes the physical ordering of the internal loss.
The characteristic-function reduction, summarized in
Appendix~\ref{app:iwop}, pulls the ideal output parity backward through the
inverse OPA, internal loss, phase shift, and first OPA. The result is an
effective single-mode observable acting on the prepared input state,
\begin{equation*}
  \langle\Pi_b\rangle_{\eta_a,\eta_b}^{(j,m)}(\phi)
  =
  \operatorname{Tr}_b\!\left[
  \rho_j^{(m)}\Omega_\phi
  \right],
\end{equation*}
The characteristic-function integral defining $\Omega_\phi$ is derived once in
Appendix~\ref{app:iwop}. The coefficients that contain the active
interferometer and the two internal loss channels are
\begin{align*}
  A_\phi
  &=
  CS
  \left(\sqrt{\eta_a}\,\ee^{\ii\phi}-\sqrt{\eta_b}\right),\\
  B_\phi
  &=
  \left(
  C^2\sqrt{\eta_b}
  -S^2\sqrt{\eta_a}\,\ee^{-\ii\phi}
  \right).
\end{align*}
\begin{align*}
  \Lambda
  &=
  (1-\eta_a)S^2+(1-\eta_b)C^2,\\
  \Gamma_\phi
  &=
  \Lambda+|A_\phi|^2,\\
  \tau_\phi
  &=
  A_\phi\alpha^* .
\end{align*}
This is the characteristic-function counterpart of the Wigner-function parity treatment used for lossy non-Gaussian phase measurements \cite{Zhang2021NCO}, now combined with the IWOP normal-ordering machinery \cite{Fan1992PRA,FanGuo2007,Fan2008IWOP}. The two-mode active interferometer and internal loss are absorbed into $A_\phi$, $B_\phi$, and $\Lambda$.

The three quantities have separate physical roles. The coefficient $A_\phi$ is the backward-propagated parity displacement into the coherent input port; after inserting $\ket{\alpha}$ it produces the displacement parameter $\tau_\phi=A_\phi\alpha^*$. The coefficient $B_\phi$ is the part of the same displacement that returns to the heralded squeezed input mode. The scalar $\Lambda$ is the Gaussian noise width accumulated from internal loss. Thus internal loss does not merely reduce a final contrast factor: it changes the operator that the prepared non-Gaussian state is tested against.

We now keep only the closed operator needed for the numerical analysis. With
\begin{equation*}
  \Delta_\phi=\Gamma_\phi+|B_\phi|^2 ,
\end{equation*}
the IWOP calculation in Appendix~\ref{app:iwop} gives the closed analytic result
\begin{widetext}
\begin{equation}
  \Omega_\phi
  =
  \frac{1}{\Delta_\phi}
  \exp\!\left[-\frac{2|\tau_\phi|^2}{\Delta_\phi}\right]
  :
  \exp\!\left[
  -\frac{2|B_\phi|^2}{\Delta_\phi}b^\dagger b
  +\frac{2B_\phi\tau_\phi}{\Delta_\phi}b^\dagger
  +\frac{2B_\phi^*\tau_\phi^*}{\Delta_\phi}b
  \right]:
  =
  \frac{1}{\Delta_\phi}
  :
  \exp\!\left[
  \frac{2(B_\phi b^\dagger-\tau_\phi^*)(\tau_\phi-B_\phi^*b)}
  {\Delta_\phi}
  \right]:
  ,
  \label{eq:Omega_normal_ordered}
\end{equation}
\end{widetext}
Equation~\eqref{eq:Omega_normal_ordered} is the closed normal-ordered effective parity observable under internal loss. The denominator $\Delta_\phi=\Gamma_\phi+|B_\phi|^2$ is the Gaussian width of the pulled-back kernel, and the exponent gives the displaced and broadened parity test seen by the prepared $b$-mode state. For $\eta_a=\eta_b=1$, it reduces at every phase to the ideal kernel $\mathcal P_\phi$ of Eq.~\eqref{eq:ideal_parity_effective_operator}; internal loss therefore deforms the same effective measurement instead of replacing it.

Only in this lossless limit, at the dark point $\phi=0$, do $A_\phi=0$, $B_\phi=1$, and $\Delta_\phi=1$, giving
\begin{equation}
  \Omega_0=:\ee^{-2b^\dagger b}:=\Pi_b .
  \label{eq:Omega_ideal_limit}
\end{equation}
Thus the effective observable reduces to ideal parity when the two OPAs cancel
and no photon is lost. The finite-derivative evaluation for PS, PA, and PC is
given in Appendix~\ref{app:iwop}; the numerical scans below insert those
derivatives into the same error-propagation formula as in Sec.~III.B. Internal
loss before the second OPA changes the effective observable itself, not merely
a final readout contrast.
\subsection{Conditional parity improvement under internal loss}

The analytic observable is now inserted in the same order as in the ideal discussion:
first the conditional parity CFI under internal loss is compared with the
Gaussian input, and only afterwards is the success probability included. The
fixed-seed parameters are $r=0.55$, $|\alpha|=1.2$, and $g=0.75$, and the
internal transmissions are $\eta_a=\eta_b=0.95$.

\begin{figure*}[t]
\includegraphics[width=\textwidth]{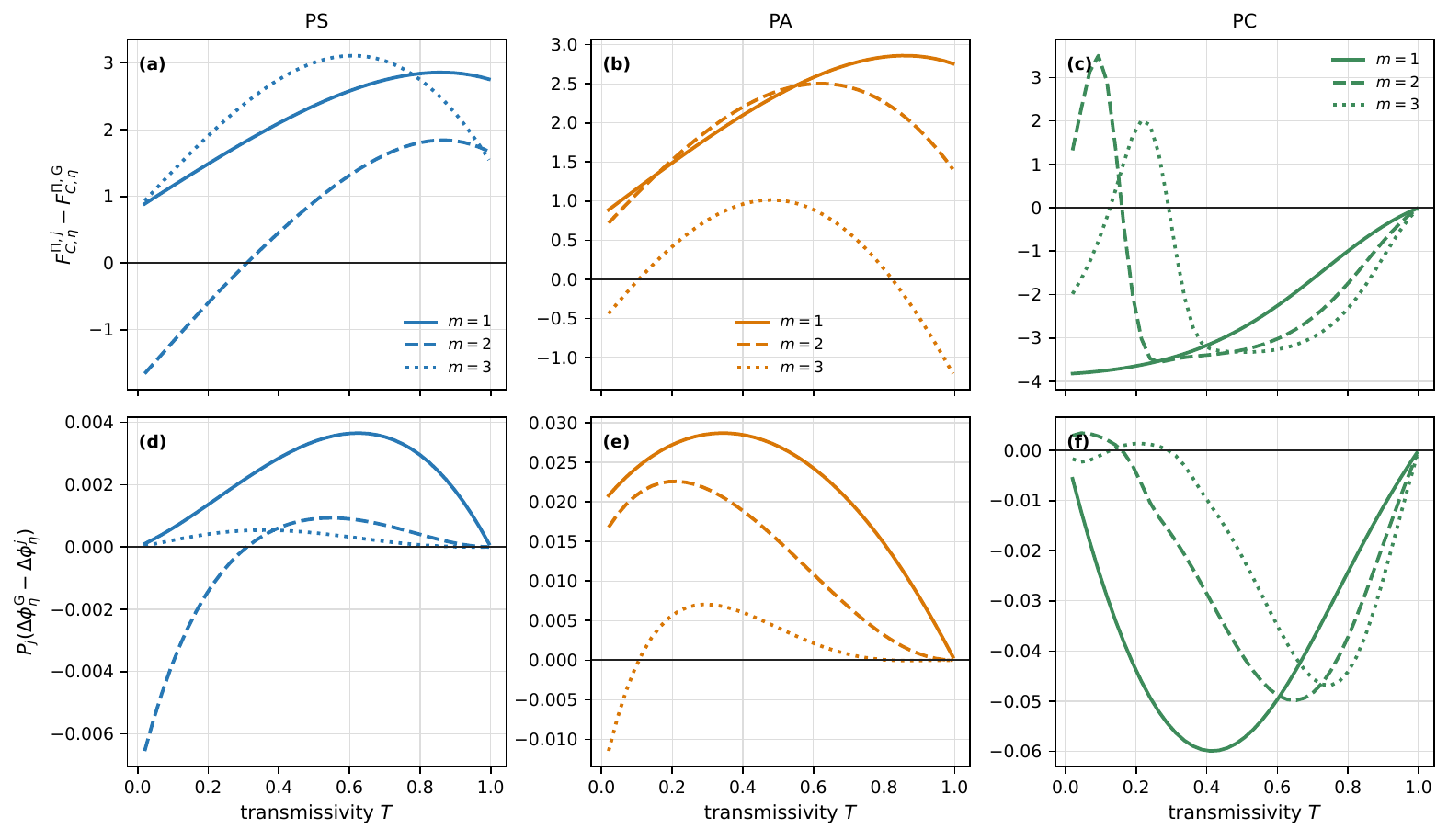}
\caption{Conditional parity performance under equal internal loss, $\eta_a=\eta_b=0.95$, with ideal output parity. The upper row gives $\max_\phi F_{C,\eta}^{\Pi,j,m}-\max_\phi F_{C,\eta}^{\Pi,\mathrm G}$, with each input optimized independently over the full phase interval. The lower row gives the diagnostic $P_j(\Delta\phi_{\eta}^{\rm G}-\Delta\phi_{\eta}^{j})$ formed from those independently optimized sensitivities. Its sign is the same as the conditional difference; weighting changes only its magnitude and operation ordering. Parameters are $r=0.55$, $|\alpha|=1.2$, and $g=0.75$.}
\label{fig:internal_loss_m123}
\end{figure*}

Figure~\ref{fig:internal_loss_m123} shows that internal loss does not remove all conditional non-Gaussian improvement, but it changes the useful transmissivity windows. For PS, the $m=1$ and $m=3$ branches are above the Gaussian noisy parity CFI throughout the displayed range, while $m=2$ becomes better only for $T\gtrsim0.31$. The largest PS conditional improvement occurs near $T=0.85$ for $m=1$, near $T=0.87$ for $m=2$, and near $T=0.61$ for $m=3$.

PA has a broader useful window: $m=1$ and $m=2$ are above the Gaussian value throughout the scan, while $m=3$ improves only in an intermediate window, approximately $0.11\lesssim T\lesssim0.82$. For PC, the single-photon branch remains below the Gaussian value, but higher-order catalysis opens low-$T$ windows: $m=2$ improves for $T\lesssim0.16$, and $m=3$ improves roughly for $0.13\lesssim T\lesssim0.29$. The noisy comparison therefore must include $m=2$ and $m=3$.

\subsection{Relative performance under internal loss}

The lower row of Fig.~\ref{fig:internal_loss_m123} applies the diagnostic of Sec.~III.C. Because the probability is positive, its zero crossings are identical to those of the conditional sensitivity difference. Weighting changes only the magnitude and the ordering: PA has the largest diagnostic value among the displayed branches, with an $m=1$ maximum of about $2.9\times10^{-2}$ at low-to-intermediate transmissivity. PS is smaller, and the positive PC values occur only for the higher-order low-transmissivity windows identified above.

This ordering is only a comparison of the probability-rescaled differences. It does not imply superiority over a fixed-gain Gaussian benchmark and does not include the Fock-ancilla cost of PA or PC. The operational per-attempt comparison is instead made with $P_jF_{C,\eta}^{\Pi,j}$ below.

\subsection{Representative fixed-resource noisy benchmarks}

The noisy parity signal obtained from $\operatorname{Tr}(\rho_j\Omega_\phi)$ is conditional on a successful heralding event. Since heralding precedes phase encoding and the internal loss channels, $P_j$ is independent of $\phi$, $\eta_a$, and $\eta_b$. The benchmark in this subsection is evaluated at selected internal transmissions and is not a full optimization over the loss map. For the representative $m=1$ resource benchmarks, we distinguish the conditional noisy parity CFI from the success-weighted information per ideal module attempt:
\begin{equation}
  F_{C,\eta,\mathrm{eff}}^{\Pi,j}
  =
  P_jF_{C,\eta}^{\Pi,j},
  \qquad
  \Delta\phi_{\Pi,\eta,\mathrm{eff}}^{(j)}
  =
  \left[P_jF_{C,\eta}^{\Pi,j}\right]^{-1/2}.
  \label{eq:loss_weighted_parity}
\end{equation}
The conditional quantity $F_{C,\eta}^{\Pi,j}$ characterizes the accepted probe under internal loss. Both $F_{C,\eta,\mathrm{eff}}^{\Pi,j}=P_jF_{C,\eta}^{\Pi,j}$ and its inverse-square-root sensitivity include failed heralding events, while remaining ideal-module metrics rather than laboratory rates.

Table~\ref{tab:loss_benchmark} gives an illustrative benchmark with equal internal transmissivity $\eta_a=\eta_b=0.95$ between the phase shift and the second OPA. The output parity measurement is ideal. The parity signal is evaluated from Eq.~\eqref{eq:Omega_normal_ordered} in a single-mode space with $d=50$; continuous bounded optimization follows a phase-grid search. The table covers representative preparation points rather than a fully optimized internal-loss map. It reports Protocols I and II only; Protocol III was used above to diagnose fixed sensing-arm exposure and is not repeated for the noisy benchmark.

This table is a representative benchmark, not a full lossy-resource optimization.

\begin{table*}[t]
\caption{Conditional and success-weighted parity CFI under equal internal loss. Here $\eta_a=\eta_b=0.95$, the output parity measurement is ideal, and the analytic $\Omega_\phi$ calculation uses $d=50$. PNR herald detection is ideal. The $F_{C,\eta}^{\Pi}$ columns condition on heralding, whereas the $P_jF_{C,\eta}^{\Pi}$ columns include failed preparations per ideal module attempt. Protocol II uses the conditional-probe slice at fixed energy and $r=0.55$; Protocol III is not included in this representative noisy table.}
\label{tab:loss_benchmark}
\begin{ruledtabular}
\begin{tabular}{lcccc}
Input & \multicolumn{2}{c}{Protocol I} & \multicolumn{2}{c}{Protocol II}\\
 & $F_{C,\eta}^{\Pi}$ & $P_jF_{C,\eta}^{\Pi}$ & $F_{C,\eta}^{\Pi}$ & $P_jF_{C,\eta}^{\Pi}$\\
\hline
Gaussian SV & $6.466$ & $6.466$ & $12.546$ & $12.546$\\
PS & $9.299$ & $0.189$ & $9.173$ & $0.187$\\
PA & $9.299$ & $0.813$ & $9.173$ & $0.802$\\
PC & $6.188$ & $5.419$ & $12.292$ & $10.766$
\end{tabular}
\end{ruledtabular}
\end{table*}

The position of internal loss matters. On the fixed-total slice, changing
$(\eta_a,\eta_b)$ from $(0.95,0.95)$ to $(0.90,0.95)$ and $(0.95,0.90)$ changes
the conditional parity CFI from $12.546$ to $7.923$ and $9.940$ for the Gaussian
input, from $9.173$ to $5.527$ and $4.523$ for PS/PA, and from $12.292$ to
$7.844$ and $10.444$ for PC. Thus loss in arm $a$ is more damaging to the
Gaussian and PC rows at these points, whereas loss in arm $b$ is more damaging
to the PS/PA row. The conditional response is input dependent; after success
weighting, the same arm dependence is additionally modulated by the
operation-specific preparation probability.

The ideal parity entries require no cutoff-dependent phase optimization because Eq.~\eqref{eq:dark_point_cfi_moments} is analytic. For the noisy calculation, increasing the single-mode cutoff from $d=30$ to $d=50$ changes every value in Table~\ref{tab:loss_benchmark} and in the asymmetric checks below the displayed precision; the $d=70$ check is supplied with the numerical data. Across representative $T=0.10$, $0.50$, and $0.90$ points for all $m=1,2,3$ curves in Fig.~\ref{fig:internal_loss_m123}, the largest absolute CFI change between $d=40$ and $d=50$ is $2.8\times10^{-9}$.

The reported phase optima for the fixed-seed rows occur at $\phi=0.14062$ (Gaussian), $0.13295$ (PS/PA), and $0.14584$ (PC); the fixed-total-slice optima are $0.11256$, $0.13344$, and $0.11533$, respectively. The values are generated from the single-mode pulled-back observable, with the ideal and $g=0$ limits checked analytically.

At $T=0.93$, Fig.~\ref{fig:internal_loss_m123} explains the entries in Table~\ref{tab:loss_benchmark}. The $m=1$ PS/PA branches still have conditional noisy parity advantage on the fixed-seed slice, whereas the $m=1$ PC branch remains slightly below the Gaussian row. After multiplying by the preparation probability, none of the listed non-Gaussian values exceeds its Gaussian counterpart. PC is nevertheless the largest of the three non-Gaussian weighted values because its preparation probability is high. This ranking is limited to the stated preparation and internal-loss parameters.

\section{Discussion}

Our results carry two main physical messages. First, PS and PA are useful as conditional filters in the high-transmissivity fixed-seed regime, especially for dark-point parity, but their per-attempt information is reduced by heralding probability and by conditional-probe resource constraints. PA gives larger probability-rescaled diagnostic values than PS because its success probability is larger, but this ranking does not include the extra cost of preparing the nonclassical Fock ancilla. Second, PC is most useful as a low-transmissivity filter and, in the enlarged fixed-gain scan, as a way to create a distinct conditional branch with high local quantum Fisher information. The Gaussian squeezed input remains the appropriate reference when one optimizes the same fixed-gain coherent--squeezed allocation without conditional filtering.

The enlarged PC scan separates encoded phase information from information accessible to dark-point parity. At its conditional-QFI optimum, the coherent amplitude is nearly zero and the number-variance term supplies most of the QFI. This term belongs to the parity-even part of the phase generator and is absent from Eq.~\eqref{eq:dark_point_cfi_moments}. The low parity-extraction ratio is therefore a measurement mismatch: the state contains phase information that the chosen binary readout does not access. This branch is a positive result of the filtering map, but accessing it efficiently would require a measurement other than dark-point single-mode parity. Because a fixed mean photon number does not bound the number variance, this high conditional QFI is a local, asymptotic quantum Cram\'er--Rao result; it does not by itself establish a finite-sample or global mean-square-error advantage over a finite prior interval \cite{Demkowicz2015,Pezze2018}.

Resource optimization also changes the interpretation of photon operations. The fixed-seed PS/PA increase results from changing the prepared-state moments at one Gaussian allocation. Once the coherent--squeezed allocation is optimized at fixed total energy and fixed gain, that increase disappears. The success-weighted PC optima occur near the identity limit of the catalysis map. Within the independently optimized success-weighted domain $0\leq r\leq1.25$, non-Gaussian filtering reshapes the conditional information but does not improve the information per ideal module attempt over the fixed-gain Gaussian optimum.

A practical caveat concerns Protocol~II. It fixes the conditional-probe energy $\bar N_{\rm enc}$ but not the average energy consumed per preparation attempt, which includes the seed squeezed-vacuum energy expended during failed heralding events. For PS and PA, whose success probabilities are of order $10^{-2}$ at the parameters of Table~\ref{tab:fixed_resource}, the average seed consumption per accepted probe is larger than $\bar N_{\rm enc}=9$ alone would suggest. Protocol~II therefore corresponds most naturally to a herald-then-store experiment in which only successfully heralded probes are injected into the interferometer \cite{Combes2014Postselection}; a fixed-rate experiment without storage would require a separate average-energy resource constraint.

This conclusion is consistent with earlier positive reports rather than contradicting them. Prior SU(1,1) studies of photon-operated inputs \cite{Gong2016Intramode,Guo2018PhotonAddedSU11,Xu2023PhotonOps,Kang2024MultiPS} use different operation locations, measurements, or resource conventions. Output-port subtraction \cite{Jiang2024OutputPS} and delocalized internal subtraction \cite{Li2025DelocalizedPS} act on states that have already passed through part of the active interferometer, while the present operations are input-side preparation filters. Similarly, homodyne-based lossy analyses for arbitrary inputs \cite{Jana2026Arbitrary} ask a different measurement question from the parity readout studied here. The IWOP result of Sec.~IV is complementary to Wigner-function parity treatments of lossy non-Gaussian phase measurements \cite{Zhang2021NCO}: it packages the active SU(1,1) transformations and internal loss into a single effective parity observable.

Input-side and internal operations remain physically distinct because $\hat K_jS_2(g)\neq S_2(g)\hat K_j$. The present herald-first ordering makes $P_j$ phase independent; an internal operation could change the phase dependence of the success probability, the noise response, and the optimal measurement.

The numerical values use moderate squeezing ($r=0.55$), high-transmissivity modules for PS/PA, and a common fixed SU(1,1) gain of $g=0.75$. At the resource level $\bar N_{\rm enc}=9$, full gain optimization would raise the Gaussian QFI to $\approx123.8$ at $g\approx1.10$; the fixed $g=0.75$ captures $\approx87\%$ of this value, so the no-advantage conclusion is conservative with respect to this restriction. The coherent-plus-squeezed-vacuum input family is the reference Gaussian probe in most SU(1,1) experiments \cite{Marino2012,Ou2012,Li2016ParitySU11}. These parameter choices are representative of a near-term optical implementation, but the ideal PNR heralding and Fock-state ancillas remain demanding. Lower internal transmission reduces the useful conditional windows; asymmetric loss can change the operation ranking because it acts differently on the coherent and heralded ports. A full loss map is a separate optimization problem.

The scope is limited to input-side operations, ideal herald detection, fixed gain, the coherent-plus-squeezed Gaussian reference family, parity readout, and selected loss points. A full operation map requires joint optimization over input allocation, gain, operation order, module transmissivities, and asymmetric loss; laboratory-rate claims additionally require ancilla generation probabilities, false heralds, mode matching, and timing.

\section{Conclusion}

We derived a unified finite-transmissivity heralding map, arbitrary-order moment generators, moment-based QFI and dark-point parity formulas, and a loss-dressed effective parity observable for input-side photon subtraction, photon addition, and photon catalysis in an SU(1,1) interferometer. The IWOP reduction of internal loss to a single pulled-back operator is exact and avoids a two-mode Fock-space propagation.

For fixed preparation parameters without resource matching, single-photon subtraction and addition improve the conditional QFI and dark-point parity information over most of the high-transmissivity regime, with stronger enhancement at higher transmissivity and at higher operation order. Multi-photon catalysis opens useful low-transmissivity conditional windows that single-photon catalysis does not provide; it can additionally generate a high-squeezing branch with large local quantum Fisher information. This branch is dominated by photon-number variance, however, and dark-point parity extracts only a small fraction of the encoded information---a measurement mismatch rather than a state-preparation failure. Under the tested internal-loss setting ($\eta_a=\eta_b=0.95$, ideal output parity), subtraction and addition retain a conditional parity advantage over the Gaussian input, while single-photon catalysis remains below the Gaussian parity benchmark.

When the coherent--squeezed allocation is independently optimized at fixed conditional-probe energy, fixed gain $g=0.75$, and $m=1$, the success-weighted Fisher information of all three non-Gaussian operations remains below the optimized Gaussian benchmark. This conclusion is subject to the tested constraints: it does not include optimization over $g$, general two-mode Gaussian probes, failed-preparation energy, ancilla-generation costs, detector inefficiency, or a full asymmetric-loss map. The three operations therefore serve distinct purposes: subtraction and addition are conditional filters for dark-point operation when a high-transmissivity module is available; catalysis is a low-transmissivity filter at $m\ge2$; and the Gaussian coherent-plus-squeezed input remains the preferred choice for per-attempt phase information under the resource constraints examined here.

\begin{acknowledgments}
This work is supported by the National Natural Science Foundation of China (Grant No.~12564049), the Jiangxi Provincial Natural Science Foundation (Grant No.~20242BAB26009), the Jiangxi Provincial Key Laboratory of Advanced Electronic Materials and Devices (Grant No.~2024SSY03011), and the Jiangxi Civil-Military Integration Research Institute (Grant No.~2024JXRH0Y07).
\end{acknowledgments}

\appendix

\section{Arbitrary-order moment generators for PS, PA, and PC}
\label{app:moments}

The single-photon PS/PA and PC moments are derived in Sec.~\ref{sec:analytic_moments}. Closed finite-differential generators for arbitrary $m$ follow from
\begin{equation}
 \begin{aligned}
 G(z)&=(1-z)^{-1/2},& z&=T^2\tanh^2 r,\\
 R&=1-T,& \mathcal D&=z\partial_z.
 \end{aligned}
 \label{eq:app_generator_definitions}
\end{equation}
together with $(x)_{\underline q}=x(x-1)\cdots(x-q+1)$ and
$(x)_{\underline0}=1$. The PS and PA success probabilities at order $m$
follow directly from their finite-transmissivity Kraus operators:
\begin{align}
  P_{\ps}^{(m)}
  &=
  \frac{R^mT^{-m}}{m!\cosh r}
  \prod_{q=0}^{m-1}(2\mathcal D-q)G(z),\nonumber\\
  P_{\pa}^{(m)}
  &=
  \frac{R^m}{m!\cosh r}
  \prod_{q=1}^{m}(2\mathcal D+q)G(z).
  \label{eq:app_m_success}
\end{align}
Define
\begin{equation}
 W_{\ps}^{(m)}(\mathcal D)=\prod_{q=0}^{m-1}(2\mathcal D-q),
 \qquad
 W_{\pa}^{(m)}(\mathcal D)=\prod_{q=1}^{m}(2\mathcal D+q).
 \label{eq:app_pspa_W}
\end{equation}
The moment-generating functions of the \emph{output} photon number are
\begin{align}
  \mathcal Z_{\ps}^{(m)}(s)
  &=
  \frac{R^mT^{-m}\ee^{-ms}}{m!\cosh r}
  W_{\ps}^{(m)}(\mathcal D_s)G(z\ee^{2s}),\nonumber\\
  \mathcal Z_{\pa}^{(m)}(s)
  &=
  \frac{R^m\ee^{ms}}{m!\cosh r}
  W_{\pa}^{(m)}(\mathcal D_s)G(z\ee^{2s}),
  \label{eq:app_m_generating}
\end{align}
where $\mathcal D_s=(z\ee^{2s})\partial_{(z\ee^{2s})}$. Hence
\begin{equation}
 \begin{aligned}
 P_j^{(m)}&=\mathcal Z_j^{(m)}(0),\\
 N_j^{(m)}&=\left.\partial_s\ln\mathcal Z_j^{(m)}(s)\right|_{s=0},\\
 V_j^{(m)}&=\left.\partial_s^2\ln\mathcal Z_j^{(m)}(s)\right|_{s=0}.
 \end{aligned}
 \label{eq:app_m_NV}
\end{equation}
Write Eq.~\eqref{eq:sv_coeff} as $\ket{\xi}=\sum_{\ell\geq0}c_\ell\ket{2\ell}$. Neighboring coefficients obey
\begin{equation}
 c_{\ell+1}\sqrt{(2\ell+2)(2\ell+1)}
 =-\ee^{\ii\theta_s}\tanh r\,(2\ell+1)c_\ell.
 \label{eq:app_neighboring_sv}
\end{equation}
Inserting the operation-dependent amplitudes on the two sides of $b^2$ then converts the remaining factor $(2\ell+1)$ into $2\mathcal D+1$. For PS and PA this gives
\begin{align}
 M_{\ps}^{(m)}
 &=-\ee^{\ii\theta_s}T\tanh r\,
 \frac{(2\mathcal D+1)W_{\ps}^{(m)}(\mathcal D)G(z)}
 {W_{\ps}^{(m)}(\mathcal D)G(z)},\nonumber\\
 M_{\pa}^{(m)}
 &=-\ee^{\ii\theta_s}T\tanh r\,
 \frac{(2\mathcal D+1)\displaystyle\prod_{q=3}^{m+2}(2\mathcal D+q)G(z)}
 {W_{\pa}^{(m)}(\mathcal D)G(z)}.
 \label{eq:app_m_pspa_M}
\end{align}
For $m=1$, Eq.~\eqref{eq:app_m_pspa_M} reduces to
$M_{\ps}=M_{\pa}=-\ee^{\ii\theta_s}3\sqrt z/(1-z)$, in agreement
with Eq.~\eqref{eq:ps_pa_moments} after the stated phase choice.

For PC, introduce the finite polynomial
\begin{equation}
 Q_m(x;T)=T^{m/2}\sum_{q=0}^{m}
 \binom{m}{q}\left(-\frac{R}{T}\right)^q
 \frac{(2x)_{\underline q}}{q!}.
 \label{eq:app_pc_Q}
\end{equation}
Equation~\eqref{eq:cat_amp} then gives
$A_{2l}^{(m)}=T^lQ_m(l;T)$. Therefore the PC normalization generator,
success probability, number moments, and pair-coherence moment are
generated using
the neighboring product $Q_m(l;T)Q_m(l+1;T)$. More explicitly, if
$G(z)=\sum_{\ell\geq0}w_\ell z^\ell$, the numerator remaining after
Eq.~\eqref{eq:app_neighboring_sv} is proportional to
\begin{equation}
 \sum_{\ell\geq0}w_\ell z^\ell(2\ell+1)
 Q_m(\ell;T)Q_m(\ell+1;T).
 \label{eq:app_pc_neighbor_sum}
\end{equation}
This sum is generated by
\begin{equation}
 \widetilde Q_m(\mathcal D;T)
 =Q_m(\mathcal D;T)Q_m(\mathcal D+1;T).
 \label{eq:app_pc_shifted_Q}
\end{equation}
Explicitly,
\begin{align}
 Z_{\pc}^{(m)}&=Q_m(\mathcal D;T)^2G(z),\nonumber\\
 P_{\pc}^{(m)}&=\frac{Z_{\pc}^{(m)}}{\cosh r},\nonumber\\
 N_{\pc}^{(m)}&=2\mathcal D\ln Z_{\pc}^{(m)},\nonumber\\
 V_{\pc}^{(m)}&=4\mathcal D^2\ln Z_{\pc}^{(m)},\nonumber\\
 M_{\pc}^{(m)}&=-\ee^{\ii\theta_s}T\tanh r\,
 \frac{(2\mathcal D+1)\widetilde Q_m(\mathcal D;T)G(z)}
 {Z_{\pc}^{(m)}}.
 \label{eq:app_m_pc_moments}
\end{align}
For $m=1$, $Q_1(\mathcal D;T)=T^{-1/2}(T-2R\mathcal D)$;
Eq.~\eqref{eq:app_m_pc_moments} is then equivalent to
Eqs.~\eqref{eq:pc_NV}--\eqref{eq:pc_M_differential}. Equations
\eqref{eq:app_m_NV}, \eqref{eq:app_m_pspa_M}, and
\eqref{eq:app_m_pc_moments} complete the theory input needed by the
order-resolved QFI and dark-point parity calculations. The exact PS--PA
coincidence in Tables~\ref{tab:mzi_standard} and \ref{tab:fixed_resource}
is recovered only at $m=1$.

\section{IWOP derivation of the ideal pulled-back parity operator}
\label{app:ideal_parity_iwop}

Section~III\,B states the reduced single-mode ideal parity kernel; this appendix collects only the IWOP evaluation of that integral. It does not introduce a second effective operator.
To evaluate the single-mode integral stated in Sec.~III\,B, use the IWOP identity
\begin{equation}
  D_b(\beta)
  =
  \ee^{-|\beta|^2/2}
  :\ee^{\beta b^\dagger-\beta^*b}: .
  \label{eq:app_ideal_displacement_iwop}
\end{equation}
With $\Delta_\phi^{(0)}=|A_\phi^{(0)}|^2+|B_\phi^{(0)}|^2$, the integral becomes
\begin{widetext}
\begin{equation}
  \mathcal P_\phi
  =
  \frac{1}{2\pi}
  :\!\int d^2\zeta\,
  \exp\!\left[
  -\frac{\Delta_\phi^{(0)}}{2}|\zeta|^2
  +\bigl(B_\phi^{(0)}b^\dagger-\tau_\phi^{(0)*}\bigr)\zeta
  +\bigl(\tau_\phi^{(0)}-B_\phi^{(0)*}b\bigr)\zeta^*
  \right]\!:
  .
  \label{eq:app_ideal_parity_iwop_integral}
\end{equation}
\end{widetext}
Within the normal-ordering symbol, $b$ and $b^\dagger$ are ordered parameters. The remaining integral is the ordinary complex Gaussian
\begin{equation}
  \int d^2z\,\ee^{-a|z|^2+bz+cz^*}
  =
  \frac{\pi}{a}\ee^{bc/a},
  \qquad \operatorname{Re}a>0,
  \label{eq:app_ideal_gaussian_integral}
\end{equation}
which directly gives Eq.~\eqref{eq:ideal_parity_effective_operator}. This completes the ideal calculation; the loss-dressed coefficients are derived separately in Appendix~\ref{app:iwop}.

\section{Characteristic-function and IWOP derivation of the noisy parity operator}
\label{app:iwop}

We give the derivation of the effective operator $\Omega_\phi$ used in the main text. For a two-mode state,
\begin{equation}
  \chi_\rho(\lambda_a,\lambda_b)
  =
  \operatorname{Tr}\!\left[
  \rho D_a(\lambda_a)D_b(\lambda_b)
  \right],
  \qquad
  D(\lambda)=\ee^{\lambda a^\dagger-\lambda^*a}.
\end{equation}
The parity expectation can be written as a phase-space integral,
\begin{equation}
  \langle\Pi_b\rangle
  =
  \frac{1}{2\pi}
  \int d^2\zeta\,\chi_\rho(0,\zeta),
  \label{eq:app_parity_characteristic}
\end{equation}
and the adjoint of a pure-loss channel acts on displacement operators as
\begin{equation}
  \mathcal L_\eta^\dagger[D(\lambda)]
  =
  \ee^{-(1-\eta)|\lambda|^2/2}D(\sqrt{\eta}\lambda).
  \label{eq:app_loss_displacement}
\end{equation}
In the main text the output parity measurement is ideal. Hence only the two
internal loss channels are pulled backward.

The remaining step is a Gaussian back-propagation of the displacement variables. From Eq.~\eqref{eq:s2_heisenberg},
\begin{equation}
  (\lambda_a,\lambda_b)
  \xrightarrow{S_2(g)}
  (C\lambda_a-S\lambda_b^*,C\lambda_b-S\lambda_a^*),
  \qquad
  \lambda_a\xrightarrow{U_\phi}\lambda_a \ee^{\ii\phi}.
\end{equation}
Starting from $(0,\zeta)$ at the detected output and pulling backward through
$S_2(-g)$, internal loss, $U_\phi$, and $S_2(g)$ gives
\begin{equation}
  (\lambda_a,\lambda_b)_{\rm in}=(A_\phi\zeta^*,B_\phi\zeta),
  \label{eq:app_backprop_variables}
\end{equation}
with $A_\phi=CS(\sqrt{\eta_a}\,\ee^{\ii\phi}-\sqrt{\eta_b})$ and $B_\phi=C^2\sqrt{\eta_b}-S^2\sqrt{\eta_a}\,\ee^{-\ii\phi}$, obtained after the successive transformations $(0,\zeta)\xrightarrow{S_2(-g)}(S\zeta^*,C\zeta)\xrightarrow{\mathcal L}(S\sqrt{\eta_a}\zeta^*,C\sqrt{\eta_b}\zeta)\xrightarrow{U_\phi}(S\sqrt{\eta_a}\ee^{\ii\phi}\zeta^*,C\sqrt{\eta_b}\zeta)\xrightarrow{S_2(g)}(A_\phi\zeta^*,B_\phi\zeta)$.
The scalar Gaussian width accumulated by internal loss is
\begin{equation}
  \Lambda
  =
  (1-\eta_a)S^2+(1-\eta_b)C^2.
\end{equation}
For the product input $\ket{\alpha}\bra{\alpha}\otimes\rho_j$, the noisy parity signal is thus
\begin{equation}
  \langle\Pi_b\rangle_{\eta_a,\eta_b}^{(j)}
  =
  \frac{1}{2\pi}\int d^2\zeta\,
  \ee^{-\Lambda|\zeta|^2/2}
  \chi_\alpha(A_\phi\zeta^*)\chi_j(B_\phi\zeta),
  \label{eq:app_noisy_signal_characteristic}
\end{equation}
where
\begin{equation}
  \chi_\alpha(\mu)
  =
  \exp\!\left[-\frac{|\mu|^2}{2}
  +\mu\alpha^*-\mu^*\alpha\right].
\end{equation}
Inserting the coherent-state characteristic function leaves an operator acting only on mode $b$,
\begin{equation}
  \Omega_\phi
  =
  \frac{1}{2\pi}
  \int d^2\zeta\,
  \ee^{-\Gamma_\phi|\zeta|^2/2+\tau_\phi\zeta^*-\tau_\phi^*\zeta}
  D_b(B_\phi\zeta),
  \label{eq:app_Omega_integral}
\end{equation}
with $\Gamma_\phi=\Lambda+|A_\phi|^2$ and $\tau_\phi=A_\phi\alpha^*$.

The remaining operator-ordering step is identical to Appendix~\ref{app:ideal_parity_iwop}: insert Eq.~\eqref{eq:app_ideal_displacement_iwop}, set $\Delta_\phi=\Gamma_\phi+|B_\phi|^2$, and apply Eq.~\eqref{eq:app_ideal_gaussian_integral}. This gives the unique normal-ordered result in Eq.~\eqref{eq:Omega_normal_ordered}; repeating the same IWOP integral here would add no new physics.

The coefficients also show the leading loss response directly: at $\phi=0$,
unequal loss produces a residual coherent-port displacement, whereas equal loss
still changes $B_0$ and $\Lambda$.

We now record the finite-derivative evaluation used for the non-Gaussian
inputs. Introduce unnormalized Bargmann kernels for mode $b$,
\begin{equation}
  \|x\rangle=\ee^{x b^\dagger}\ket{0},
  \qquad
  \langle y\|=\bra{0}\ee^{y b},
  \label{eq:bargmann_kernel}
\end{equation}
With the phase-matched convention $\theta_s=\pi$, the squeezed vacuum takes the Bargmann form
$\ket{\xi}=(\cosh r)^{-1/2}\exp(\frac{\lambda}{2}b^{\dagger 2})\ket{0}$ with $\lambda=\tanh r$.
Equation~\eqref{eq:Omega_normal_ordered} gives
\begin{equation}
  \mathcal E_\phi(x,y)
  =
  \langle y\|\Omega_\phi\|x\rangle
  =
  \kappa_\phi
  \exp\!\left[
  \sigma_\phi yx+\ell_\phi y+\nu_\phi x
  \right],
  \label{eq:Omega_bargmann_kernel}
\end{equation}
with $\kappa_\phi=\Delta_\phi^{-1}\exp[-2|\tau_\phi|^2/\Delta_\phi]$, $\sigma_\phi=1-2|B_\phi|^2/\Delta_\phi$, $\ell_\phi=2B_\phi\tau_\phi/\Delta_\phi$, and $\nu_\phi=2B_\phi^*\tau_\phi^*/\Delta_\phi$.
For any Gaussian kernel $\mathcal F(x,y)=\kappa\exp(a yx+\ell y+\nu x)$, the squeezed-vacuum contraction is
\begin{equation}
  \mathfrak S_r[\mathcal F]=\bra{\xi}\mathcal F\ket{\xi}
  =\frac{\kappa}{\cosh r\sqrt{1-\lambda^2a^2}}
  \exp\!\left[\frac{\lambda(\nu^2+\ell^2+2\lambda a\ell\nu)}{2(1-\lambda^2a^2)}\right].
  \label{eq:squeezed_kernel_contraction}
\end{equation}

For the unnormalized state
$\tilde\rho_j^{(m)}=K_j^{(m)}\ket{\xi}\bra{\xi}K_j^{(m)\dagger}$, define
\begin{align}
  \widetilde\Pi_j^{(m)}(\phi)
  &=
  \operatorname{Tr}(\tilde\rho_j^{(m)}\Omega_\phi),
  &
  P_j^{(m)}&=\operatorname{Tr}\tilde\rho_j^{(m)},
  \nonumber\\
  \langle\Pi_b\rangle^{(j,m)}_\eta
  &=
  \frac{\widetilde\Pi_j^{(m)}(\phi)}{P_j^{(m)}}.
  \label{eq:normalized_parity_from_tilde}
\end{align}
For PS,
\begin{equation}
  \widetilde\Pi_{\ps}^{(m)}(\phi)=\frac{R^m}{m!}\,\partial_u^m\partial_v^m\,
  \mathfrak S_r\!\bigl[\ee^{ux+vy}\mathcal E_\phi(\sqrt{T}x,\sqrt{T}y)\bigr]\big|_{u=v=0},
  \label{eq:ps_parity_diff_closed}
\end{equation}
and for PA,
\begin{equation}
  \widetilde\Pi_{\pa}^{(m)}(\phi)=\frac{R^m}{m!}\,\partial_u^m\partial_v^m\,
  \mathfrak S_r\!\bigl[\mathcal E_\phi(\sqrt{T}x+u,\sqrt{T}y+v)\bigr]\big|_{u=v=0}.
  \label{eq:pa_parity_diff_closed}
\end{equation}
For PC, the beam-splitter action on Bargmann sources gives
\begin{equation}
  B_{bc}(T)\|x\rangle_b\|u\rangle_c
  =
  \|\sqrt{T}x-\sqrt{R}u\rangle_b
  \|\sqrt{R}x+\sqrt{T}u\rangle_c ,
  \label{eq:bs_bargmann_sources}
\end{equation}
and hence
\begin{equation}
  \begin{aligned}
  \widetilde\Pi_{\pc}^{(m)}(\phi)&=\frac{1}{(m!)^2}\,
  \partial_u^m\partial_v^m\partial_s^m\partial_t^m\\
  &\times\mathfrak S_r\!\bigl[\ee^{t(\sqrt{R}x+\sqrt{T}u)}\ee^{s(\sqrt{R}y+\sqrt{T}v)}\\
  &\times\mathcal E_\phi(\sqrt{T}x-\sqrt{R}u,\sqrt{T}y-\sqrt{R}v)\bigr]\big|_{u=v=s=t=0}.
  \end{aligned}
  \label{eq:pc_parity_diff_closed}
\end{equation}
The success probabilities follow by replacing
$\mathcal E_\phi(x,y)$ with the identity kernel
\begin{equation}
  \mathcal E_\phi(x,y)\rightarrow \langle y\|x\rangle=\ee^{yx}.
  \label{eq:success_from_identity_kernel}
\end{equation}
Since $P_j^{(m)}$ is independent of $\phi$ for the input-side operations,
\begin{equation}
  \partial_\phi
  \langle\Pi_b\rangle_{\eta_a,\eta_b}^{(j,m)}
  =
  \frac{\partial_\phi\widetilde\Pi_j^{(m)}(\phi)}{P_j^{(m)}}.
  \label{eq:diff_parity_derivative}
\end{equation}

\end{document}